\documentclass[lettersize,journal]{IEEEtran}
\usepackage{amsmath,amsfonts}
\usepackage[normalem]{ulem}
\usepackage{algorithmic}
\usepackage{algorithm}
\usepackage{array}
\usepackage[caption=false,font=normalsize,labelfont=sf,textfont=sf]{subfig}
\usepackage{textcomp}
\usepackage{stfloats}
\usepackage{url}
\usepackage{verbatim}
\usepackage{graphicx}
\usepackage{cite}
\usepackage{xcolor}
\usepackage{orcidlink}
\usepackage{soul}
\begin{document}

\definecolor{dynamiccolor}{RGB}{102, 194, 165}
\definecolor{sparsecolor}{RGB}{141, 160, 203}
\definecolor{ourscolor}{RGB}{252, 141, 98}


\newcommand{\remove}[1]{}
\newcommand\addition[1]{{\color{black}#1}}
\newcommand\additiontwo[1]{{\color{black}#1}}
\newcommand{\removetwo}[1]{}

\newcommand{\stkout}[1]{\ifmmode\text{\sout{\ensuremath{#1}}}\else\sout{#1}\fi}


\title{NeRF-CA: Dynamic Reconstruction of X-ray Coronary Angiography with Extremely Sparse-views }

\author{
Kirsten W.H. Maas~\orcidlink{0009-0007-4402-0301}, 
Danny Ruijters~\orcidlink{0000-0002-9931-4047}, 
Anna Vilanova~\orcidlink{0000-0002-1034-737X}~\IEEEmembership{,~Member,~IEEE}, 
and Nicola Pezzotti~\orcidlink{0000-0001-9554-4331}
\thanks{Kirsten W.H. Maas and Anna Vilanova are with the Department of Mathematics \& Computer Science, Eindhoven University of Technology, 5612AZ Eindhoven, The Netherlands (e-mail: k.w.h.maas@tue.nl; a.vilanova@tue.nl).

Danny Ruijters is with Philips Healthcare, 5684PC Best, The Netherlands,
and also with the Department of Electrical Engineering, Eindhoven University of Technology, 5612AZ Eindhoven, The Netherlands
(e-mail: danny.ruijters@philips.com).

Nicola Pezzotti was with the Department of Mathematics \& Computer Science,
Eindhoven University of Technology, 5612AZ Eindhoven, The Netherlands,
and also with Philips Healthcare, 5684PC Best, The Netherlands.
He is now with ASML, 5504DR Veldhoven, The Netherlands
(e-mail: nicola.pezzotti@asml.com).}
}



\maketitle

\begin{abstract}
\addition{
Dynamic three-dimensional (4D) reconstruction from two-dimensional X-ray coronary angiography (CA) remains a significant clinical problem. 
Existing CA reconstruction methods often require extensive user interaction or large training datasets. 
Recently, Neural Radiance Field (NeRF) has successfully reconstructed high-fidelity scenes in natural and medical contexts without these requirements. 
However, challenges such as sparse-views, intra-scan motion, and complex vessel morphology hinder its direct application to CA data. 
We introduce NeRF-CA, a first step toward a fully automatic 4D CA reconstruction that achieves reconstructions from sparse coronary angiograms. 
To the best of our knowledge, we are the first to address the challenges of sparse-views and cardiac motion by decoupling the scene into the moving coronary artery and the static background, effectively translating the problem of motion into a strength. 
NeRF-CA serves as a first stepping stone for solving the 4D CA reconstruction problem, achieving adequate 4D reconstructions from as few as four angiograms, as required by clinical practice, while significantly outperforming state-of-the-art sparse-view X-ray NeRF.
We validate our approach quantitatively and qualitatively using representative 4D phantom datasets and ablation studies. 
To accelerate research in this domain, we made our codebase public: \url{https://github.com/kirstenmaas/NeRF-CA}.
}

\end{abstract}

\begin{IEEEkeywords}
4D reconstruction, X-ray coronary angiography, Neural Radiance Field, sparse-view, static and dynamic decomposition
\end{IEEEkeywords}

\section{Introduction}


X-rays are fundamentally limited by providing a two-dimensional (2D) representation of a three-dimensional (3D) structure, resulting in a lack of depth perception. 
This problem is especially apparent in X-ray coronary angiography (CA) interventions, where dynamic sequences of 2D X-ray projections or coronary angiograms represent complex dynamic 3D (4D) vessel structures.
On top of that, in standard CA interventions, an extremely limited set of these angiogram views are acquired per patient, which we refer to as sparse-view \cite{green2004three}.
These interventions could benefit from 4D reconstructions to enhance the 4D perception of the vessel structure for catheter navigation applications \cite{piayda2018dynamic}.
Nonetheless, accurate 4D reconstruction from CA data poses a challenge due to inherent characteristics from the X-ray system, complex vessel characteristics, and intra-scan motion \cite{ccimen2016reconstruction}. 
The challenges of the X-ray system include extremely sparse-view X-ray projections and system geometry inaccuracies. 
The vessel characteristics include structure sparsity, vessel overlap, and \remove{background occlusion due to overlapping background structures or contrast inhomogeneity}\addition{poor vessel visibility due to background obstruction, related to overlapping background structures or blood vessel contrast inhomogeneity}.  
Moreover, coronary angiography data exhibit intra-scan motion due to cardiac and respiratory motion.
\remove{Although clinically relevant}\addition{Overall}, 4D reconstruction of CA data remains a challenge.

Several CA reconstruction methods have been proposed, including traditional and machine learning methods~\cite{ccimen2016reconstruction, iyer2023multi, zhu2025sparse}.
Both methods can obtain satisfactory reconstructions from extremely sparse-view sequences.
\remove{However, they rely on time-consuming user segmentations, error-prone background removal techniques, or large amounts of training data.}
\addition{However, they rely on time-consuming user segmentations or large amounts of training data, consisting of many views per patient, which are not available due to the clinical sparse-view setting of CA.}
These dependencies hinder the adaptation of these methods in clinical practice.
While some methods ignore the dynamics, others assume cyclic cardiac motion through an acquired electrocardiogram (ECG) signal but still lead to motion artifacts \cite{ccimen2016reconstruction}.

Neural Radiance Field (NeRF) is a promising deep learning technique for 3D reconstruction that can reconstruct high-fidelity 3D natural scenes from 2D projections \cite{mildenhall2021nerf}.
Unlike machine learning methods that learn a prior, NeRF learns an individual scene given a set of 2D images at hand.
As such, it does not rely on segmentations or large amounts of training data, showing the potential to overcome the limitations of prior CA reconstruction techniques.
Recent work explored the potential of NeRF in X-ray angiography, showing that limitations still lie in reconstruction in a sparse-view setting~\cite{maas2023nerf}.

Meanwhile, many extended NeRF works have been proposed to reconstruct in the sparse-view setting \cite{gao2022nerf}. 
Regularization-based techniques for natural scenes have been shown to reconstruct from only $3$ views \cite{yang2023freenerf}. 
\addition{
Techniques designed for natural scenes rely on properties such as depth and sparsity, which are absent in the CA setting, making them unsuitable.}
Concurrently, NeRF-based techniques proposed for medical scenes do not reach the extremely sparse view requirements for CA scenes, requiring a minimum of 15 projections for adequate performance~\cite{cai2024structure}. 
\remove{Furthermore, methods for reconstructing video-acquired natural dynamic scenes have also been \mbox{proposed~\cite{gao2022monocular}.}}
\addition{
NeRF methods, such as 3DGR-CAR~\cite{fu20243dgr} and NeCA~\cite{wang2024neca}, address sparse-views for CA scenes. 
However, they rely on accurate segmentation or extensive training data and ignore the inherent motion of CA data.}
\addition{On the other hand, }Liu et al.~\cite{liu20243ddsa} recently presented static and dynamic scene decomposition for modeling vessel dynamics in digital subtraction angiography (DSA) separating foreground and background.
Although similar to CA, DSA does not suffer from \remove{background occlusion}\addition{poor vessel visibility}, and the technique proposed by Liu et al.~\cite{liu20243ddsa} is not adequate for sparse-view settings since it requires many training projections.

We propose NeRF-CA, the first step towards a 4D X-ray Coronary Angiography (CA) reconstruction method addressing \addition{the challenges of} sparse-views and the cardiac motion characteristics of CA acquisitions.
We present 4D vessel reconstructions from an extremely sparse amount of \addition{synthetic} coronary angiogram sequences exhibiting cardiac motion.
We \remove{leverage}\addition{build on} the extended work of NeRF for static and dynamic natural scene decomposition \cite{wu2022d} and techniques for reconstruction in a sparse-view setting \cite{yang2023freenerf, kim2022infonerf}.
Specifically, we utilize the rapid motion of the coronary artery to decouple the scene into a dynamic coronary artery component and a static background component.
\remove{To perform the dynamic reconstruction, we uniquely combine scene decomposition with a weighted pixel sampling technique, static and dynamic factorization regularization, and entropy regularization, enforcing an accurate foreground and background separation.
For sparse-view reconstruction, we combine these techniques with coarse-to-fine positional encoding and occlusion regularization.}
\addition{The novelty of our work lies in developing a combination of self-supervised decomposition with regularization techniques to enable sparse-view reconstructions for CA scenes.
Specifically, we propose regularization techniques for the dynamic coronary artery component, enforcing properties such as scene sparsity and minimal occlusion.}
Together, we demonstrate vessel reconstruction capabilities from as few as four coronary angiogram \remove{sequences}\addition{views}, matched to clinical workflows.
We validate the \remove{relevance}\addition{effect} of these constraints through ablation studies.
\addition{Similar to other related works~\cite{iyer2023multi, fu20243dgr},} we demonstrate our method on two 4D phantom datasets~\cite{segars20104d, rosset2004osirix}, simulating synthetic coronary angiograms.
\addition{
These synthetic coronary angiograms allow us to address the challenges of sparse-views and cardiac motion in CA data, having a ground-truth as realistic as possible.
}
\remove{This synthetic imaging also allows us to} \addition{Meanwhile, we can} perform the necessary quantitative validation, as we can acquire ground-truth sequences from any viewpoint.
We compare our method with X-ray sparse-view NeRF reconstruction methods, significantly outperforming them in the extremely sparse-view X-ray coronary angiography setting.
\addition{This provides the initial validation of the method in an ideal setting, showing the potential for future developments and applications to real data.}

\remove{In summary, the main contribution of our work is a novel method for 4D reconstruction of extremely sparse-view  CA data addressing cardiac motion.}
\addition{In summary, the main contribution is a methodology to address cardiac motion and sparse-view characteristics for 4D reconstruction of CA data.
This serves as a first step towards a solution for clinical data.}
Unlike previous methods, our method does not rely on manual user interactions or large datasets.
We decouple the coronary angiography scene in a dynamic coronary artery and static background by uniquely combining scene decomposition with regularization constraints fitting the sparse nature of the blood vessel structure. 
Moreover, we perform experiments to compare our method to \remove{other}\addition{existing state-of-the-art NeRF-based} methods while validating the relevance of our choices through ablation studies.

\section{Background}
In this section, we will discuss the clinical background of coronary X-ray angiography and the technical background of Neural Radiance Fields (NeRFs), which are relevant to understanding our work.

\subsection{X-ray Coronary Angiography} \label{sec:cag}
X-ray coronary angiography (CA) is the most common imaging modality to diagnose and treat coronary artery disease.
Although it has fundamental limitations due to its 2D representation of a 3D anatomy, it still outperforms 3D imaging modalities.
The benefits are its \remove{usage for diagnosis and treatment and} spatial and temporal resolution \cite{chen2009three}.
This work focuses on CA imaging acquired by a single-plane C-arm angiography system, which is still the gold standard for CA interventions \cite{knuuti20202019}.

\begin{figure}[t]
     \centering
     \includegraphics[width=\linewidth]{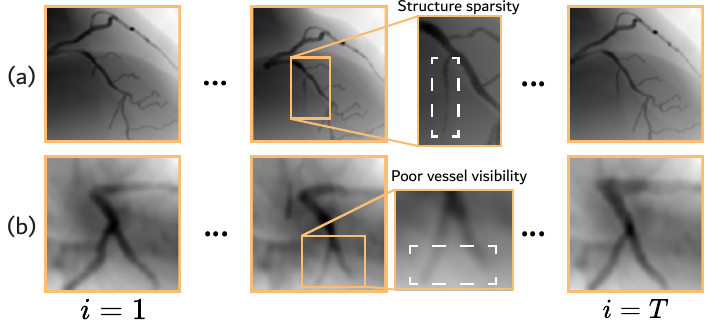}
    \caption{One coronary angiogram sequence for the (a) XCAT \cite{segars20104d} and (b) MAGIX \cite{rosset2004osirix} datasets with cardiac phases $i \in \{ 1, \dots, T \}$.}
    \label{fig:examples}
\end{figure}

A coronary angiogram sequence consists of multiple frames showing a contrast-enhanced coronary artery experiencing cardiac and respiratory motion.
\addition{In single-plane CA, these sequences are obtained from a single acquisition angle or view. 
The terms "sequence" and "view" will be used interchangeably throughout this paper, depending on whether the focus is on the temporal or directional aspect, respectively.}
\remove{\mbox{Figure~\ref{fig:examples}} provides an example of one coronary angiogram sequence generated from two synthetic datasets across multiple cardiac phases $i$ within one cardiac cycle.}
\addition{In Figure~\ref{fig:examples}, an example of one coronary angiogram sequence is shown for two datasets, where each frame represents a single cardiac phase $i$ within one cardiac cycle.}
As can be seen, the imaging contains sparse blood vessel structures that may overlap\addition{, as highlighted in Figure~\ref{fig:examples}~(a)}.
\addition{Moreover, the imaging suffers from poor vessel visibility through overlapping background structures or inhomogeneous contrast, which is highlighted in Figure~\ref{fig:examples}~(b). }
In a clinical setting, an extremely sparse amount of angiogram \remove{sequences} \addition{views} (i.e., 2-4 per patient) is captured to minimize X-ray exposure.
C-arm systems can be rotated using 3D Euler angles $\theta$ and $\phi$ to obtain varying viewing points \cite{ccimen2016reconstruction}.
\remove{The operator chooses these viewing points, typically based on standard optimal viewing points described in \mbox{literature~\cite{di2005coronary}.}}
The coronary arteries experience cardiac and respiratory motion while acquiring the angiogram sequences.
\addition{Cardiac motion dominates the dynamics of the coronary arteries~\cite{shechter2006displacement}.}
\remove{The cardiac motion of the coronary arteries}\addition{This motion} aligns with the beating\remove{ motion} of the heart, which is assumed to be synchronous with the simultaneously recorded electrocardiogram (ECG).
Therefore, individual frames of the coronary angiogram can be assigned an estimated cardiac phase within each cardiac cycle, as shown in Figure~\ref{fig:examples}.
\remove{Cardiac motion dominates the \mbox{dynamics~\cite{shechter2006displacement}.}}
\remove{However, respiratory motion also occurs, which is correlated with breathing rate.}
\addition{On the other hand, respiratory motion is correlated with the breathing rate, which is typically not measured clinically.
As a result, this motion is more complex to estimate~\cite{shechter2006displacement} and is assumed to be residual for 3D reconstruction tasks~\cite{ccimen2016reconstruction}.}

4D reconstructions in the context of CA can aid both the diagnosis and treatment of coronary artery disease. 
\remove{The 2D CA sequences may lead to a suboptimal assessment of stenosis severity and, therefore, suboptimal stent size selection, which a 4D reconstruction could overcome. 
Coronary roadmaps are commonly generated from a reconstruction to aid catheter navigation through blood \mbox{vessels~\cite{piayda2018dynamic}.}
The reconstruction quality for this roadmap application is typically evaluated based on the vessel topology rather than the vessel diameter \mbox{accuracy~\cite{maas2023nerf}}.}
\addition{An example of a clinical application of 4D reconstructions is called coronary roadmap, in which the reconstruction is overlayed over the X-ray to aid catheter navigation through the blood vessels~\cite{piayda2018dynamic}.
This overlay is created by re-projecting the reconstruction based on the C-arm angle of interest, similar to the novel view application of NeRFs.}


\subsection{Neural Radiance Field}
A Neural Radiance Field (NeRF) represents a single scene through a multilayer perceptron (MLP), a fully connected neural network \cite{mildenhall2021nerf}.
Given a 3D coordinate $\boldsymbol{x}=(x,y,z)$ and a viewing direction $\boldsymbol{d} = (\theta, \phi)$, the network outputs a density $\sigma$ and color $\boldsymbol{c} = (r,g,b)$ to represent natural scenes.
NeRF therefore learns a continuous function $F_\Theta(\boldsymbol{x}, \boldsymbol{d}) = (\sigma, \boldsymbol{c})$, where $\Theta$ represents the learned weights of the MLP.
\addition{A NeRF needs to be re-optimized for each scene. }

As MLPs have difficulty learning high frequencies, NeRF employs positional encoding $\gamma$ on the input parameters to allow for high-frequency details in the reconstructions \cite{mildenhall2021nerf}.
Early works used sinusoidal functions with different frequencies to map the input to a higher dimensional space \cite{tancik2020fourier}. 

NeRF renders pixels through the learned density $\sigma$ and color $\boldsymbol{c}$ by discretizing the volume rendering equation \cite{max1995optical}.
Every pixel is associated with a ray $\boldsymbol{r}$, which is defined as $\boldsymbol{r}(t) = \boldsymbol{o} + t\boldsymbol{d}$, with camera origin $\boldsymbol{o}$, ray direction $\boldsymbol{d}$ and $t$ the \remove{distance from the origin}\addition{position along the ray}.
The obtained predicted pixel color $\hat{C}(\boldsymbol{r})$ is compared to the ground truth pixel color $C(\boldsymbol{r})$ of a given image to optimize the model utilizing mean squared error (MSE) \addition{through the photometric loss}.
We refer to the work of Mildenhall et al. \cite{mildenhall2021nerf} for further details on the method.

\section{Related work}
This section will discuss X-ray coronary angiography (CA) reconstruction and Neural Radiance Field techniques related to the work in this paper.

\subsection{X-ray coronary angiography reconstruction techniques} \label{subsec:works_cag}
A wide range of X-ray coronary angiography (CA) reconstruction techniques have been proposed,
which can be categorized into model-based, tomographic-based, and machine learning-based \cite{ccimen2016reconstruction, iyer2023multi}.
\remove{We will discuss these techniques, their background removal strategies, and their strategies to handle cardiac motion.}
\addition{We will discuss these techniques and their strategies to handle cardiac motion and sparse-views.}

Model-based techniques extract the vessels from the CA imaging to reconstruct them as binary trees.
These methods achieve accurate reconstruction with sparse clinical views but rely on time-consuming manual input~\cite{chen2009three}.
Recently, machine learning approaches have been proposed for automatic segmentation~\cite{hwang2021simple, bappy2021automated} or direct 3D reconstruction~\cite{iyer2023multi, zhao2022self, zhu2025sparse}. 
\remove{However, these methods require extensive segmented training data, often unavailable in coronary angiography.}
\addition{However, these methods require large amounts of accurately segmented training data, including multiple views per patient, which is unfeasible in clinical settings due to the need to minimize X-ray exposure and the time-consuming nature of segmentation.}
\additiontwo{
Although automatically obtained segmentations offer a potential solution, 3D reconstruction methods that depend on them are highly sensitive to segmentation quality. 
As stated by several works~\cite{hwang2021simple, bappy2021automated, zhu2025sparse}, inaccuracies, such as missing vessels or inconsistencies across 2D views, can propagate through the reconstruction process, resulting in noisy or incomplete 3D reconstructions.}
Tomographic techniques reconstruct attenuation volumes from raw X-ray imaging without needing manual segmentation. 
However, they require many views, limiting their use in single-plane CA.
\remove{Methods based on NeRF could potentially overcome these limitations, which will be discussed in \mbox{Section~\ref{subsec:works_nerf}}.}
\addition{Methods based on NeRF have the potential to overcome these limitations, and will be discussed in Section~\ref{subsec:works_nerf}.}
Our work will build on these methods in the context of CA.

\remove{CA imaging captures both coronary arteries and background structures, which can cause truncation errors due to incomplete visibility from every viewpoint. 
As a result, background removal through suppression or subtraction is a common technique applied in tomographic reconstruction. 
As a pre-processing step, good suppression and subtractions can be achieved, but they are hindered by the dynamics of the CA imaging, leading to background motion \mbox{artifacts~\cite{rohkohl2010interventional}}.
Other works perform background subtraction during reconstruction, which have shown to be effective in separating the vessel from the \mbox{background~\cite{liu2014improved}}.
Inspired by these methods, our work also separates the blood vessels from the background to avoid truncation errors by decoupling the dynamic coronary artery from the static background.}

Traditional 3D reconstruction methods use various strategies to mitigate motion artifacts, primarily focusing on cardiac motion and treating respiratory motion as residual~\cite{ccimen2016reconstruction}. 
\addition{For the cardiac motion}, they use ECG signals for frame extraction through gating or motion compensation. 
Gating selects frames around the same cardiac phase, while motion compensation aligns all frames to the same phase. 

\remove{Our work focuses on cardiac motion, assuming
residual respiratory motion.}
\addition{Our work focuses on solving the challenge of cardiac motion within the sparse-view setting, not considering respiratory motion and other real-world data aspects, like noise or patient motion.
NeRF-CA serves as a first step towards a solution for the full clinical problem, given that a method has yet to be proposed that addresses cardiac motion and sparse-views without the reliance on manual segmentations or large datasets, where both factors have so far hindered the development of a clinically applicable solution to the reconstruction problem.} 
Incorporating respiratory motion \addition{and other real-world data aspects} \remove{should be explored in future work} \addition{is out of scope for this paper, such that the cardiac motion and sparse-view challenges can be isolated and evaluated independently.
In case of potential results, future work should address the remaining data aspects.}

\subsection{Neural Radiance Fields} \label{subsec:works_nerf}

NeRFs have been widely adopted for reconstructing natural scenes due to their ability to reconstruct high-fidelity \remove{scenes}\addition{details}. 
However, challenges persist in reconstructing natural scenes from sparse-views, especially when dynamic~\cite{yunus2024recent}. 
On top of that, CA scenes present unique challenges such as sparse structures and \remove{background occlusion}\addition{poor vessel visibility}~\cite{wang2024neural, molaei2023implicit}. 

Many works have addressed the sparse-view problem, which can be categorized into prior-based and regularization-based methods \cite{gao2022nerf}.
Whereas prior-based methods train a general model on large datasets, regularization-based methods introduce additional scene-specific regularization terms. 
Given the lack of data in single-plane CA, as discussed in Section~\ref{subsec:works_cag}, prior-based methods are unsuitable in our context.
Regularization-based methods instead rely on scene-specific properties \cite{niemeyer2022regnerf, jain2021putting, kim2022infonerf, yang2023freenerf}.
However, these methods are yet to be adapted to medical scenes, which pose unique properties such as \remove{background occlusion}\addition{poor vessel visibility and therefore complicating foreground and background distinction}.
For medical scenes, reconstruction in a sparse-view setting also poses a significant challenge \cite{ruckert2022neat, zang2021intratomo}. 
Prior-based \cite{lin2023learning, fang2022snaf, kshirsagar2024generative} or X-ray regularization-based \cite{cai2024structure, zha2022naf, zang2021intratomo} strategies have been proposed to address this in static scenes, reaching a limit in performance with $15$ projections \cite{cai2024structure}.
However, this number of projections is still inadequate for our clinical context and does not account for sparse structures in dynamic scenes. 
\remove{In our work, we leverage the regularization-based methods proposed for natural scenes by adapting them to the blood vessel properties.}
\addition{In our work, we are inspired by the regularization-based methods utilized for natural scenes.
We propose regularization-based methods that fit blood vessel properties, effectively translating regularization-based techniques to the medical imaging setting.}

\remove{On the other hand, many strategies have been proposed for the 4D reconstruction of dynamic \mbox{scenes~\cite{gao2022monocular}}.
For medical scenes, some works have been proposed to model the dynamics, but they require a significantly large amount of \mbox{projections~\cite{zang2021intratomo, zhang2023dynamic, birklein2023neural}}.}
NeRF methods have also been proposed for blood vessel reconstruction, addressing sparse-views and dynamics \addition{independently}.
\addition{For CA scenes, several NeRF-based techniques have been proposed, including 3DGR-CAR~\cite{fu20243dgr} and NeCA~\cite{wang2024neca}. 
These methods address the sparse-view problem, enabling reconstructions from as few as two projections. 
However, they depend on accurate segmentation or extensive training data and overlook the inherent dynamics of CA data.}
For vessels that are not prone to extensive motion, digital subtraction angiography (DSA) can be used to remove all background structures from the 2D projection images.
TiaVox \cite{zhou2023tiavox} and Liu et al. \cite{liu20243ddsa} successfully reconstruct these sparse structure scenes from only $30$ DSA projections.
Liu et al. \cite{liu20243ddsa} model blood flow dynamics by decoupling the scene into a static background and dynamic blood vessel foreground.
Static and dynamic scene decomposition is one of the approaches to model dynamics with NeRFs, where scenes can be decoupled in a static and dynamic component \cite{yunus2024recent}. 
However, these DSA scenes do not experience \remove{background occlusion}\addition{poor vessel visibility}, recently highlighted as a significant challenge in sparse-view X-ray angiography reconstruction with NeRFs~\cite{maas2023nerf}.
Moreover, CA scenes exhibit even more significant dynamics in foreground and background than DSA scenes. 

In this work, we decouple the static and dynamic scenes and extend this strategy to address the properties of CA scenes.
\addition{To the best of our knowledge, NeRF-CA is the first to effectively address the challenges of sparse-views and dynamics of CA data.}
\remove{We build on a natural scene decomposition technique \mbox{D$^2$NeRF~\cite{wu2022d}}, which we adjust to CA scenes.}
\addition{Inspired by the natural scene decomposition technique D$^2$NeRF~\cite{wu2022d}, we decouple the CA scene into a dynamic coronary artery component and a background component.}
\remove{On top of that, we impose regularization on the individual decoupled scenes, including structure \mbox{sparsity~\cite{kim2022infonerf}},scene \mbox{smoothness~\cite{yang2023freenerf}}, and edge occlusion
\mbox{minimization~\cite{yang2023freenerf}}.}
\addition{We uniquely leverage this static and dynamic decomposition by imposing regularization on each learned decoupled component.
This regularization is inspired by successful techniques from the natural imaging domain, including structure sparsity~\cite{kim2022infonerf}, scene smoothness~\cite{yang2023freenerf}, and edge occlusion minimization~\cite{yang2023freenerf}.
However, their extension to the CA setting is not straightforward, hence a new methodology is needed.}
\remove{These approaches are the bases of NeRF-CA, which allows for 4D reconstruction from an extremely sparse number of angiogram sequences.}
\addition{NeRF-CA allows 4D reconstruction from an extremely sparse number of angiogram views without user interaction, which is currently not possible with state-of-the-art methods.}

\section{NeRF-CA} \label{sec:method}

\begin{figure*}[t]
    \centering
    \includegraphics[width=\textwidth]{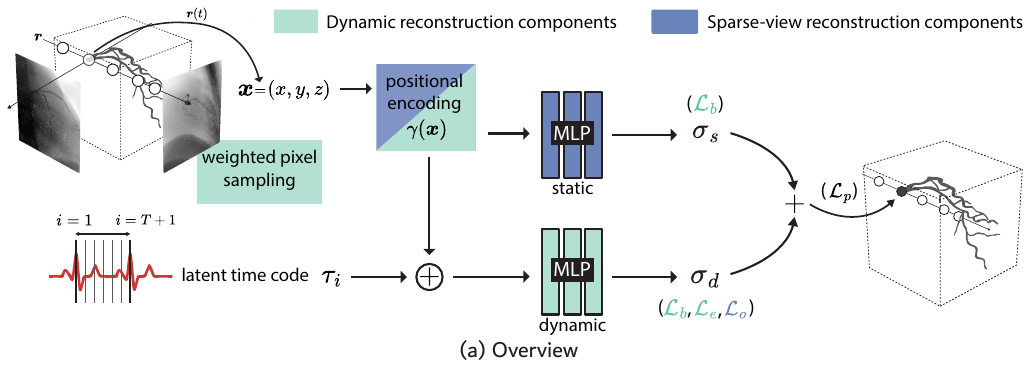}
    \includegraphics[width=\textwidth]{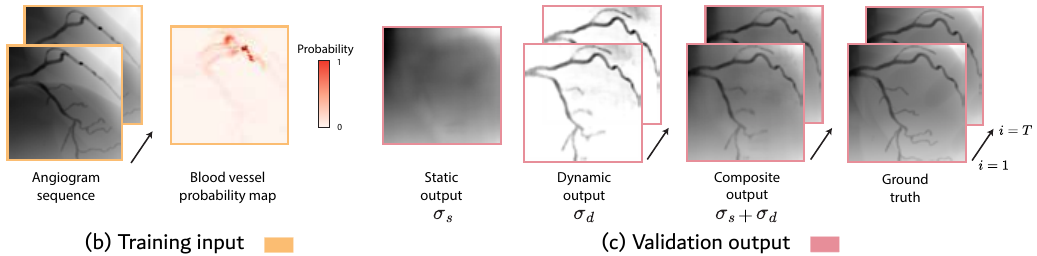}
    \caption{Overview of the NeRF-CA method (a). The full loss function $\mathcal{L}_f$ (\ref{eq:full_loss}) consists of (\ref{eq:loss_intensity}) Photometric Loss $\mathcal{L}_p$, (\ref{eq:sd_factorization}) Static vs. Dynamic Factorization Loss $\mathcal{L}_b$, (\ref{eq:d_entropy}) Dynamic Entropy Loss $\mathcal{L}_e$, and (\ref{eq:d_occlusion}) Dynamic Occlusion Loss $\mathcal{L}_o$. We also depict the (b) training input and (c) validation output, in yellow and pink, respectively. }
    \label{fig:overview}
\end{figure*}

In this section, we describe our method, NeRF-CA\remove{, which allows for dynamic Coronary Angiography reconstruction with extremely sparse-views}.
Figure~\ref{fig:overview}~(a) shows an overview of our method. 
We categorize the components of our method according to their function.
Specifically, we \remove{utilize}\addition{propose} dynamic reconstruction components, shown in green in Figure~\ref{fig:overview}~(a), and sparse-view reconstruction components, shown in purple in Figure~\ref{fig:overview}~(a).
Overall, the X-ray scene is expressed as an absorption-only model with attenuation coefficients $\sigma$, as will be further described in Section~\ref{subsec:nerfforxray}.
Figure~\ref{fig:overview}~(b, c) depicts the model training input and validation output in yellow and pink, respectively.
We will utilize these colors throughout the paper to distinguish the training \addition{(yellow)} and validation \addition{(pink)} views.

The basis of our method is decoupling the scene into a static and dynamic scene.
The static model represents the background, and the dynamic model represents the dynamic coronary artery structure. 
The static and dynamic models predict their attenuation coefficients $\sigma_s$ and $\sigma_d$, respectively.
These coefficients are composited to obtain the scene from which the predicted pixel intensities are calculated.
Figure~\ref{fig:overview}~(c) depicts the outputs of our methods, specifically, the static, dynamic, and composite outputs for one validation view.
Whereas the static model solely relies on the positionally-encoded 3D coordinate as input, the dynamic models also receive a per-cardiac phase latent code $\tau_i$.
This latent code represents a cardiac phase along the cardiac cycle derived from the simultaneously recorded ECG signal.
More details on the scene decomposition can be found in Section~\ref{subsec:decouple}.

By decomposing the scene, we can individually impose regularizers for the static and dynamic scenes, allowing the final dynamic scene to model the coronary artery.
In Section~\ref{subsec:objective}, we provide the overall objective \addition{and loss function} of our method \addition{while} introducing these regularizers.
Sections~\ref{subsec:dynamicrecon} and \ref{subsec:sparserecon} describe the regularizers for dynamic and sparse-view reconstruction, respectively.
\remove{The positional encoding of our method improves both dynamic and sparse-view reconstruction, indicated by the two colors in \mbox{Figure~\ref{fig:overview}}.}
\remove{We also perform weighted pixel sampling to guide the model to distinguish the sparse blood vessel structure from the background.
We perform this sampling with a blood vessel probability map, representing the pixel probability of having a blood vessel.
}
\addition{
A core aspect of our approach is weighted pixel sampling, which guides the model to distinguish the sparse blood vessel structures from the background. 
This is achieved using a blood vessel probability map, which estimates the likelihood of each pixel containing a blood vessel. 
}
This probability map is shown in Figure~\ref{fig:overview}~(b) and detailed in Section~\ref{subsec:dynamicrecon}.

\subsection{Overall Objective} \label{subsec:objective}
\remove{Our overall objective consists of two parts: dynamic and sparse-view reconstruction.}
\remove{Our contribution lies in adapting and combining these dynamic and sparse-view reconstruction strategies.}
\addition{Our contribution lies in developing a unique combination of dynamic and sparse-view reconstruction, enabled by an ECG-timed temporal integration of the cardiac sequence and a combination of several regularization techniques targeted to tackle the unique challenges presented by CA data.}
\remove{To demonstrate these strategies, we provide a running example throughout the explanation of our method.}
Figure~\ref{fig:overview}~(c) shows examples of the predicted coronary angiogram views and the ground truth views. 

\addition{The full loss of NeRF-CA $\mathcal{L}_f$ can be summarized as

\begin{multline} \label{eq:full_loss}
   \mathcal{L}_f(\boldsymbol{r}, \boldsymbol{\tau}_i) = \mathcal{L}_p(\boldsymbol{r}, \boldsymbol{\tau}_i) + \lambda_b\mathcal{L}_b(\boldsymbol{r}, \boldsymbol{\tau}_i) \\ + \lambda_e\mathcal{L}_e(\boldsymbol{r}, \boldsymbol{\tau}_i) + \lambda_o\mathcal{L}_o(\boldsymbol{r}, \boldsymbol{\tau}_i),
\end{multline}
where $\mathcal{L}_p$ represents the photometric loss, $\mathcal{L}_b$ represents the static vs. dynamic factorization loss with weight $\lambda_b$, $\mathcal{L}_e$ represents the dynamic entropy loss with weight $\lambda_e$, and $\mathcal{L}_o$ represents the dynamic occlusion loss with weight $\lambda_o$.
Each of these components will be introduced in the next paragraphs.
The final model optimization is performed by sampling rays $\boldsymbol{r}$ from a total set of rays $\mathcal{R}$ defined by the ground truth pixels.
Each pixel has a cardiac phase $i$ assigned, which is denoted by a time latent code $\boldsymbol{\tau}_i$, which will be detailed in Section~\ref{subsec:decouple}.
}

For dynamic reconstruction, we decouple the static background from the dynamic coronary artery structure, as detailed in Section~\ref{subsec:dynamicrecon}.
We combine \remove{standard} static and dynamic decomposition with a weighted pixel sampling strategy and two regularizers to accurately decouple the coronary artery from the background.
\addition{Our method is self-supervised through the standard photometric loss $\mathcal{L}_p$, computed based on the composition of the static and dynamic components.}
Weighted pixel sampling is \remove{introduced}\addition{proposed} to separate the sparse blood vessel structure from the background at the image level.
The static vs. dynamic factorization regularizer $\mathcal{L}_b$ is enforced to separate the static and dynamic scene on a 3D point level \cite{wu2022d}.
In addition, we enforce entropy $\mathcal{L}_e$ in the decoupled dynamic scene to achieve the sparse structure of the blood vessel \cite{kim2022infonerf}.

For sparse-view reconstruction, we reconstruct from an extremely sparse number of angiogram \remove{sequences}\addition{views}.
The sparse-view reconstruction components, consisting of windowed positional encoding and dynamic occlusion loss, will be detailed in Section~\ref{subsec:sparserecon}.
Windowed positional encoding is a proven strategy to enforce smoothness\remove{ while combating degenerate high-frequency solutions}~\cite{yang2023freenerf}.
\remove{We enforce this strategy in both the static and dynamic scenes.}
\addition{We employ this strategy in the decoupled static and dynamic scenes to enforce smoothness in both scenes, avoiding degenerate high-frequency solutions.}
The occlusion loss $\mathcal{L}_o$ punishes high occlusion areas close to the camera, an expected faulty behavior when applying NeRFs for sparse-view reconstruction~\cite{yang2023freenerf}.
We demonstrate that imposing such a loss can benefit our sparse dynamic scene.

\subsection{Neural Radiance Field for X-ray} \label{subsec:nerfforxray}
X-ray scenes can be described by the absorption-only model, assuming the Beer-Lambert law \cite{max1995optical}.
Similarly to other works applying NeRFs for X-ray scenes (e.g., \cite{ruckert2022neat}), we approximate the scene by $F_\Theta(\boldsymbol{x}) = \sigma$, where $\sigma$ represents the attenuation coefficient in the X-ray scene.
The predicted pixel intensity $\hat{I}(\boldsymbol{r})$ of an X-ray image is then described \remove{by the Beer-Lambert law}\addition{as}:
\begin{equation*}
     \hat{I}(\boldsymbol{r}) = I_0 \exp\Bigr(-\int_{t_n}^{t_f} \sigma(\boldsymbol{r}(t)) dt\Bigr),
\end{equation*}
where \addition{$t$ represents the position along the ray, and} $t_n$ and $t_f$ represent the near and far thresholds along the ray.
The X-ray source is assumed to emit a constant initial intensity $I_0$ independent of the viewing direction. 

\subsection{Static and Dynamic Scene Decomposition} \label{subsec:decouple}
The base of our method is the static and dynamic decoupling of the scene, inspired by the work of D$^2$NeRF and Liu et al. \cite{wu2022d, liu20243ddsa}\addition{, and CA reconstruction works that separate the blood vessel from the background during reconstruction~\cite{liu2014improved}}. 
Specifically, we decouple our scene in a static component $F_{\Theta_s}$ and dynamic component $G_{\Theta_d}$. Two MLPs represent them with weights $\Theta_s$ and $\Theta_d$, respectively, as shown in Figure~\ref{fig:overview}~(a).
\begin{equation*}
    F_{\Theta_s}(\gamma(\boldsymbol{x})) = \sigma_s
\end{equation*}
\begin{equation*}
    G_{\Theta_d}(\gamma(\boldsymbol{x}),  \boldsymbol{\tau}_i ) = \sigma_d
\end{equation*}

The static component $F_{\Theta_s}$ predicts a static attenuation coefficient $\sigma_s$ from a positionally-encoded 3D coordinate $\gamma(\boldsymbol{x})$, similar to the originally proposed NeRF \cite{mildenhall2021nerf}. 
We use a coarse-to-fine windowed positional encoding $\gamma$ in our work, which will be further described in Section~\ref{subsec:sparserecon}.
Unlike the static component, the dynamic component $G_{\Theta_d}$ is also conditioned on a latent code per cardiac phase $\tau_i$ as input and predicts a dynamic attenuation coefficient $\sigma_d$.
This latent code $\tau_i$ is a learned vector representing a discrete cardiac phase $i$.
Specifically, each cardiac cycle, represented by a fixed interval of the ECG as shown in Figure~\ref{fig:overview}~(a), is divided into $T$ equal parts.
The parameter $T$, where $i \in \{ 1, \dots, T \}$ \addition{$\bmod$ $T$}, is pre-defined based on the frame rate of the angiogram sequence.
In other words, we assume that the frame time matches the cardiac phase $i$.
\remove{
Note that, whereas \mbox{D$^2$NeRF~\cite{wu2022d} and Liu et al.~\cite{liu20243ddsa}} represent $\tau_i$ as per-frame time, we represent $\tau_i$ as a cardiac phase in the cardiac cycle.
}
\addition{
A key difference in our approach lies in how we define $\tau_i$.
While \mbox{D$^2$NeRF~\cite{wu2022d} and Liu et al.~\cite{liu20243ddsa}} represent $\tau_i$ as per-frame time, we instead represent $\tau_i$ as a cardiac phase in the cycle.
}
This approach allows us to match cardiac phases across viewpoints.

The static and dynamic attenuation coefficient for each 3D coordinate $\boldsymbol{x}$ are composited as $\sigma = \sigma_s + \sigma_d$ to obtain the predicted pixel intensity $\hat{I}(\boldsymbol{r}, \boldsymbol{\tau}_i)$ at cardiac phase $i$
\begin{equation*}
     \hat{I}(\boldsymbol{r}, \boldsymbol{\tau}_i) = I_0 \exp\Bigr(-\int_{t_n}^{t_f} (\sigma_s(\boldsymbol{r}(t) + \sigma_d(\boldsymbol{r}(t), \boldsymbol{\tau}_i))dt\Bigr).
\end{equation*}

The photometric loss $\mathcal{L}_p$ is computed from the predicted pixel intensity and the ground truth pixel intensity.
This photometric loss $\mathcal{L}_p$ ensures that the predicted composite frame at cardiac phase $i$ is similar to the ground truth frame at cardiac phase $i$ from the angiogram sequence.
Specifically, for every ray $\boldsymbol{r}$ at cardiac phase $i$, we optimize
\begin{equation} \label{eq:loss_intensity}
    \mathcal{L}_p(\boldsymbol{r}, \boldsymbol{\tau}_i) = 
        ( \hat{I}(\textbf{r}, \boldsymbol{\tau}_i)-I(\textbf{r}, \boldsymbol{\tau}_i) )^2.
\end{equation}

\subsection{Dynamic Reconstruction Components} \label{subsec:dynamicrecon}
The function of the static and dynamic decoupling in our work is to separate the scene in the background and the dynamic coronary artery \addition{in a self-supervised manner}.
However, the coronary artery is not the only dynamic element in our CA scene.
For example, the heart and ribs also experience motion.
Moreover, the sparsity of the structure of the blood vessel also poses a challenge in the separation.
Specifically, we propose weighted pixel sampling to address the blood vessel's separation from the image's background.
We adopt static vs. dynamic factorization loss to correctly separate static and dynamic scenes on a position level \cite{wu2022d}.
Lastly, to force our dynamic scene to only model the blood vessel, we impose a sparsity structure constraint on our dynamic scene \cite{kim2022infonerf}.
\addition{An alternative to our self-supervised approach is to directly supervise the dynamic output of our model using a coronary artery segmentation mask, such as one generated by a Frangi filter~\cite{frangi1998multiscale}, which we will ablate in \mbox{Section~\ref{subsec:ablation}}.}
Figure~\ref{fig:overview}~(c) shows an example of our dynamic reconstruction.
As can be seen, the dynamic output mostly consists of the coronary artery structure, while the static output represents the background.
Next, we will provide more details on the strategies to achieve these results.\\


\noindent \textbf{Weighted Pixel Sampling} \hspace{4pt} \remove{To optimize our model, we sample rays from all angiogram sequence frames, as described in \mbox{Section~\ref{subsec:decouple}}.}
\addition{To optimize a NeRF model, rays are sampled for every pixel of every frame, as was described in Section~\ref{subsec:decouple}.}
However, the blood vessel structure occupies only a few pixels in every coronary angiogram frame.
When all pixels are sampled uniformly, this structure may be missed and, therefore, difficult to separate from the background.
\remove{We perform weighted pixel sampling to reconstruct the blood vessels rather than the background.}
\addition{Therefore, we uniquely introduce weighted pixel sampling to reconstruct the blood vessels rather than the background.}
The weights are computed based on the assumption that the blood vessel structure is of high contrast and motion.
We calculate a sampling probability map based on the variance of each pixel value across the frames of one angiogram sequence.
Figure~\ref{fig:overview}~(b) provides an example of a probability map for one angiogram sequence.
The blood vessel structure is distinguished from the background with higher sampling probability, indicated by the darker red areas.
During training, we sample a pre-defined fraction of rays $V$ from this map's thresholded set of high-variance pixels.
The remaining pixels are sampled randomly from all possible pixels. \\


\noindent \textbf{Static vs. Dynamic Factorization Loss} \hspace{4pt} A position in space should be occupied by the static or dynamic scene but never by both. 
The photometric loss, detailed in Equation~\ref{eq:loss_intensity}, does not guarantee this behavior and, therefore, a correct separation of static and dynamic scenes.
To enforce this separation, we adopt the Static vs. Dynamic Factorization regularizer\remove{ from D$^2$NeRF}~\cite{wu2022d}.
This regularizer defines a spatial ratio of dynamic and static attenuation coefficients as
\begin{equation*}
    w(\boldsymbol{x}, \boldsymbol{\tau}_i) = \frac{\sigma_d(\boldsymbol{x}, \boldsymbol{\tau}_i)}{\sigma_d(\boldsymbol{x}, \boldsymbol{\tau}_i) + \sigma_s(\boldsymbol{x})}.
\end{equation*}
This ratio is penalized when deviating from a categorical distribution $\{0, 1\}$ through a binary entropy loss $\mathcal{L}_b$:
\begin{equation} \label{eq:sd_factorization}
    \mathcal{L}_b(\boldsymbol{r}, \boldsymbol{\tau}_i) = \int_{t_n}^{t_f} H_b(w(\boldsymbol{r}(t), \boldsymbol{\tau}_i) dt 
\end{equation}
\begin{equation*}
    H_b(x) = -(x \cdot \log(x) + (1 - x) \cdot \log(1 - x))
\end{equation*}

\remove{D$^2$NeRF also proposes a skewness on top of the binary entropy loss to slightly favor static explanations of the scene, as scenes are naturally more static than dynamic.
In our scene, the dynamic component is sparse in structure and occupies only a few points in the 3D space.
The skewness measure is generically applied to the 3D space and may lead to a degenerate solution where the dynamic component is completely removed.
We, therefore, do not incorporate this skewness measure.}
\addition{
D$^2$NeRF uses a skewness term on top of the binary entropy loss to favor static explanations, assuming scenes are mostly static. 
However, since our dynamic component is sparse and occupies only a few points in the 3D space, applying skewness leads to degenerate solutions where the dynamic component is entirely removed.
We omit the skewness term and instead achieve structure sparsity uniquely through a dynamic entropy loss, as detailed below.
} \\

\noindent \textbf{Dynamic Entropy Loss} \hspace{4pt} In the final reconstruction, the part of the scene occupied by the coronary artery should be small.
Previous works for the reconstruction of CA scenes have accomplished this by enforcing a small number of points of the scene to have non-zero attenuation values \cite{li2004improved}.
However, we would also like to force a sparsity constraint on our blood vessel structure.
We utilize ray entropy to enforce sparsity specifically on the dynamic component of our scene.
As a result, our composite scene favors most points to be static while modeling the dynamic component as a sparse structure, accomplishing an ideal static vs. dynamic factorization.
Previous works have defined ray entropy to avoid cloud-like artifacts in NeRF scenes \cite{kim2022infonerf, wu2022d}.
Based on these works, we define the ray density $ p_{\boldsymbol{r}, \boldsymbol{\tau}_i}$ as
\begin{equation*}
    p_{\boldsymbol{r}, \boldsymbol{\tau}_i}(t) = \frac{\sigma_d(\boldsymbol{r}(t), \boldsymbol{\tau}_i)}{\int_{t_n}^{t_f}\sigma_d(\boldsymbol{r}(s), \boldsymbol{\tau}_i) ds}.
\end{equation*}
The dynamic entropy loss function $\mathcal{L}_e$ is then defined as
\begin{equation} \label{eq:d_entropy}
    \mathcal{L}_e(\boldsymbol{r}, \boldsymbol{\tau}_i) = -\int_{t_n}^{t_f} p_{\boldsymbol{r}, \boldsymbol{\tau}_i}(t) \log(p_{\boldsymbol{r}, \boldsymbol{\tau}_i}(t)) dt.
\end{equation}

To avoid artifacts from appearing in empty spaces, we disregard non-hitting rays based on their accumulated density, similar to previous work \cite{kim2022infonerf}.
\remove{However, we always compute the entropy for rays likely to hit a blood vessel to avoid these rays nearing zero.}
\addition{However, given that we want to achieve sparseness for the areas that have blood vessels, we always compute entropy for rays likely to hit a blood vessel, avoiding these rays to near zero.}
We derive this likelihood from the blood vessel probability map, which we also utilize for weighted pixel sampling as described in the respective paragraph and shown in Figure~\ref{fig:overview}~(b).

\subsection{Sparse-view Reconstruction Components} \label{subsec:sparserecon}
The main goal of our work is to reconstruct a coronary artery structure from an extremely sparse amount of views.
While the dynamic reconstruction components provide a decoupled reconstruction of the coronary artery, degenerate solutions may still occur when limiting the number of views for training.
We, therefore, introduce two components \remove{inspired by existing literature} for sparse-view reconstruction\remove{ with NeRFs in natural scenes \mbox{\cite{yang2023freenerf}}}.
Namely, we incorporate windowed positional encoding $\gamma(\textbf{x})$ to achieve a smooth CA scene.
Moreover, we also impose edge occlusion regularization on the dynamic scene to avoid common degenerate solutions where artifacts appear near the camera.\\

\noindent \textbf{Windowed Positional Encoding} \hspace{4pt} A key element in reconstructing high-fidelity scenes with NeRFs is the positional encoding \cite{tancik2020fourier}.
While these resulting high-frequency inputs allow for fast reconstruction of high-frequency components, they are prone to bias the model to generate high-frequency artifacts, especially in a sparse-view setting~\cite{yang2023freenerf}. 
Several works in NeRF have introduced a frequency regularization method, which linearly increases the frequency of the positional encoding over time \cite{lin2021barf, park2021nerfies, yang2023freenerf}\addition{, enforcing smoothness in the scene}.
\remove{This method, which is called windowed positional encoding, forces a coarse-to-fine transition during training.
This method reduces high-frequency \mbox{artifacts~\cite{park2021hypernerf, park2021nerfies}}, also in the case of static and dynamic scene decomposition to enforce smooth \mbox{separation~\cite{wu2022d}}.}
Recently, FreeNeRF \cite{yang2023freenerf} has shown that their windowed positional encoding is especially effective in avoiding degenerate solutions in an extremely sparse-view setting.
\addition{This windowed positional encoding is applied on both the static and dynamic component of our scene, which enforces smoothness in each decoupled component.}
The positional encoding is\remove{ then} defined as 
\begin{equation} \label{eq:pos_enc}
    \gamma_\alpha(\boldsymbol{x}) = (\boldsymbol{x}, \dots, \alpha_L(n)\sin(2^{L}\pi\boldsymbol{x}), \alpha_L(n)\cos(2^{L}\pi\boldsymbol{x})).
\end{equation}
The weight $\alpha_L(n) = clamp(\frac{nL}{N}, 0, 1)$ is set based on the current training iteration $n$, and the hyperparameter $N$ which denotes the number of iterations needed to reach the total number of frequency bands $L$. \\

\noindent \textbf{Dynamic Occlusion Loss} \hspace{4pt} While windowed positional encoding is effective, this strategy still allows for degenerate solutions in the sparse-view setting.
Specifically, FreeNeRF~\cite{yang2023freenerf} demonstrated that NeRF models may generate a solution where high-frequency artifacts appear near the camera when presented with an extremely sparse amount of training views.
These solutions are correct for the training views but do not correctly generalize to new viewpoints.
To avoid this behavior, FreeNeRF~\cite{yang2023freenerf} proposes an occlusion regularization loss, punishing high occlusion areas close to the camera.
While this loss may be effective in natural scenes with objects in the center, our composite X-ray scene consists of a few non-zero attenuation points.
\remove{On the other hand, our dynamic scene should be close to empty, with the coronary artery close to the center due to the X-ray system setup.
Therefore, we propose to adopt the occlusion regularization loss on our dynamic scene solely.
Besides solving the problem of high occlusion areas near the camera, we also expect to achieve solutions where the dynamic scene is nearly empty together with the dynamic entropy loss enforcing sparsity.}
\addition{
As a result, applying this loss directly to the composite scene would not avoid degenerate outcomes.
In contrast, our learned dynamic scene, representing the coronary artery, is expected to be nearly empty, with the coronary artery positioned near the center due to the X-ray system setup. 
Therefore, on top of our self-supervised decomposition approach, we apply occlusion regularization exclusively to the learned dynamic scene. }
\remove{\mbox{Figure~REMOVED} shows examples of scenes trained with $4$ angiogram sequence, without dynamic occlusion loss $\mathcal{L}_o$ and our full $\mathcal{L}_f$ method.
As can be seen, the occlusion regularization loss forces the dynamic scene to be nearly empty while still modeling the sparse coronary artery structure correctly.}


We define the occlusion regularization loss $\mathcal{L}_o$ as
\begin{equation} \label{eq:d_occlusion}
    \mathcal{L}_o(\boldsymbol{r}, \boldsymbol{\tau}_i) = \int_{t_n}^{t_f} \Bigr(\sigma_d(\boldsymbol{r}(t), \boldsymbol{\tau}_i) \cdot M(t)\Bigr) dt,
\end{equation}
where $M(t)$ is a binary mask applied respectively to the distance $t$ of a point along the ray starting from the X-ray source. 
This mask evaluates to $1$ if the point's distance $t$ is lower than the distance threshold $D$, otherwise it evaluates to $0$. 
Note that FreeNeRF~\cite{yang2023freenerf} does not penalize based on a distance $t$ but rather on the first $C$ points starting from the X-ray source.
We rather penalize based on a distance threshold $D$, as this is more applicable to a C-arm system setup where physical distances between the X-ray source and patient are \remove{measured}\addition{known}. 
Specifically, we enforce the points with a distance smaller than $D$ to be outside the patient and, therefore, nearly empty in density, which matches the goal of the occlusion regularization.
\\\\ \addition{Equation~\ref{eq:full_loss} provides an overview of NeRF-CA's full loss function.}
Rather than utilizing constant weights for \addition{the hyperparameters} $\lambda_b$, $\lambda_e$, and $\lambda_o$, we linearly increase them over time with a pre-determined number of delay steps.
This approach is fitting due to the coarse-to-fine transition of the windowed positional encoding, as imposing weights too early in a low-frequency setting may lead to degenerate solutions.
 


\remove{where $\lambda_b$, $\lambda_e$, and $\lambda_o$ represent the weights for the static vs. dynamic factorization, dynamic entropy, and dynamic occlusion loss, respectively.}
\remove{The final model optimization is performed by sampling these rays $\boldsymbol{r}$ from the total set of rays $\mathcal{R}$ defined by the given ground truth pixels at cardiac phases $i$.}

\section{Experiments}
In this section, we present experiments to evaluate and validate NeRF-CA.
We present the 4D phantom datasets that allow us to evaluate quantitatively.
Our method has multiple components and parameters that we evaluate through extensive ablation studies. 
Furthermore, we compare our method to existing state-of-the-art \addition{methods} to show the strengths and weaknesses of our approach.

\subsection{Datasets}
For the evaluation, we \remove{rely on}\addition{use} parameterized 4D synthetic phantoms previously used in a similar evaluation context~\cite{rohkohl2010cavarev}. 
These phantoms allow us to generate realistic ground-truth X-ray sequences from the anatomical cardiac vessel structure from any viewpoint at any cardiac \remove{motion} phase.
\remove{Utilizing these phantoms, we can effectively focus our evaluation on the aspects NeRF-CA addresses, i.e., vessel sparsity, background occlusion, and the modeling of synchronized cardiac motion.}
\addition{Utilizing these phantoms, we can effectively focus our evaluation on the aspects NeRF-CA specifically addresses, i.e., sparse-views and cardiac motion.
Meanwhile, these phantoms also exhibit varying aspects of vessel sparsity and poor vessel visibility, allowing us to validate our model under these conditions.}
\remove{We note that these phantoms do not exhibit other complex characteristics that are present in real clinical CA data, such as blood vessel contrast inhomogeneity, lack of cardiac synchronization among views, and respiratory motion, which are not addressed by NeRF-CA.
Clinical CA data exhibiting all the complex characteristics, as described in \mbox{Section~\ref{sec:cag}}, would limit the value of the evaluation through the confounding factors influencing the results. 
Furthermore, the flexibility for ablation studies is rather limited on real clinical CA data, for example, we would like to utilize more CA sequences than the maximum of $4$ acquired per patient in the clinical setting.
Moreover, no 3D or 4D ground-truth volumes are available in clinical CA data that from which we can generate or evaluate 2D CA sequences.}
\addition{These phantoms intentionally exclude complex characteristics of real clinical CA data, such as blood vessel contrast inhomogeneity, lack of cardiac synchronization among views, and respiratory motion, to ensure the evaluation remains focused on the aspects NeRF-CA is designed to address. 
Using clinical data with these factors, as described in \mbox{Section~\ref{sec:cag}}, would introduce confounding variables, limiting the evaluation's value.
Additionally, real clinical CA data offers limited flexibility for ablation studies; for example, the typical clinical setting provides a maximum of 4 views per patient. 
Moreover, clinical CA data lacks 3D or 4D ground-truth volumes needed for generating or evaluating 2D CA sequences.
}

\remove{Therefore, }We use the 4D XCAT phantom \cite{segars20104d} and a 4D coronary computed tomography angiography (CCTA) scan, called MAGIX \cite{rosset2004osirix}.
Both datasets depict the left coronary artery (LCA), which is more complex in geometry than the right coronary artery and, therefore, more clinically interesting for 3D reconstruction applications.
For both datasets, we simulate homogeneous contrast injection using higher attenuation values in segmented areas, similar to previous work \cite{maas2023nerf}.
Figure~\ref{fig:examples} shows examples of one angiogram sequence for these 4D datasets.
In these datasets, one angiogram consists of $T = 10$ frames, each corresponding to a discrete cardiac phase $i$, assuming minimal respiratory motion.
As can be seen from Figure~\ref{fig:examples}, the 4D XCAT dataset \cite{segars20104d} includes a detailed LCA structure, fitting for evaluating the effect of vessel sparsity on our reconstruction quality.
Moreover, the 4D XCAT phantom allows for the simulation of cardiac motion according to the ECG-based cardiac phase.
We sample the XCAT phantom with an isotropic resolution of 0.5 mm$^3$.
The 4D CCTA dataset MAGIX, obtained from the OsiriX platform \cite{rosset2004osirix}, consists of 10 consecutive scans that represent different cardiac phases across the cardiac cycle.
As contrast is injected throughout the body during CCTA scans, each scan also includes varying contrast in the background, which allows us to evaluate the effect of the dynamic background on our reconstructions.
Due to the CCTA's lower resolution of $0.4\times0.4\times2$ mm$^3$, the obtained LCA structure is coarse.
To generate projections from the 4D volumes, we utilize the tomographic toolbox TIGRE \cite{biguri2016tigre}.
We adjust the geometry settings to simulate the C-arm setting of CA data, where the focus is on the main branches while deeper branches may be out of view.

\begin{figure}
    \centering
    \includegraphics[width=\linewidth]{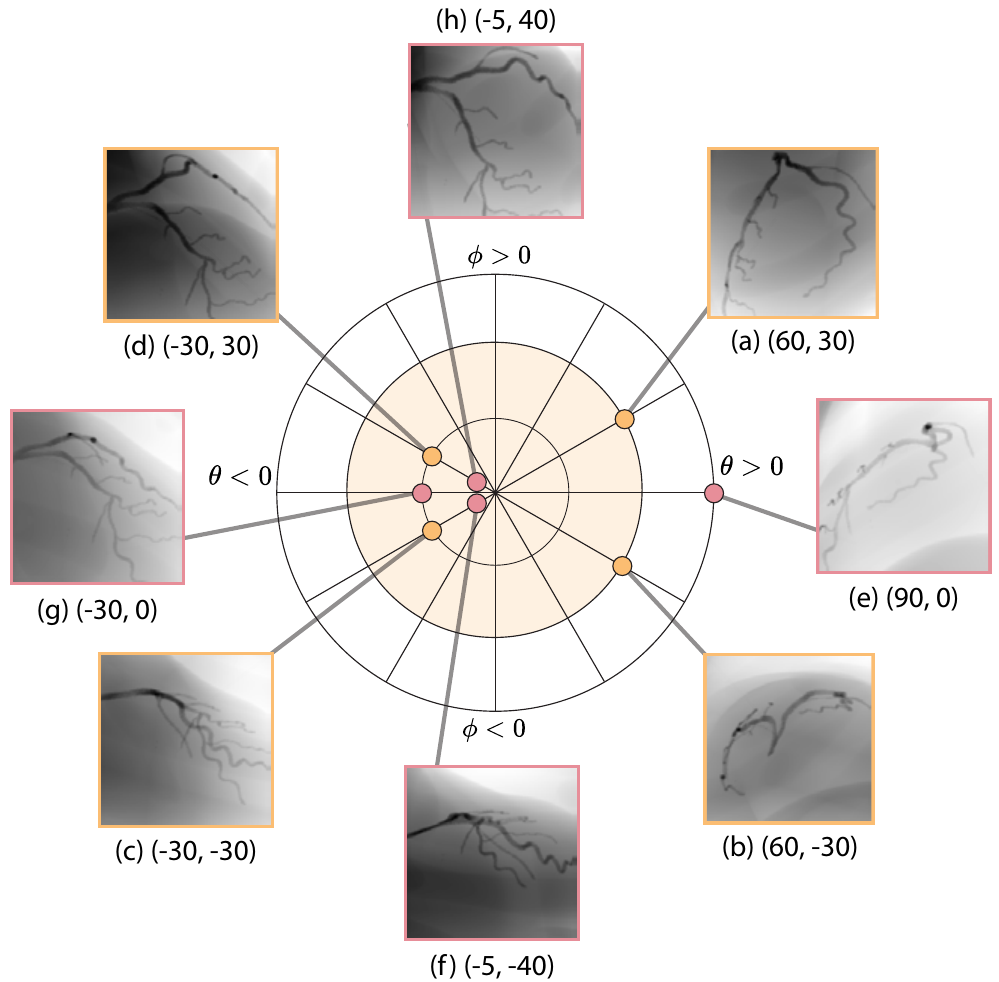}
    \caption{Optimal viewing angles for the left coronary artery expressed in Euler angles ($\theta$, $\phi$). The angles are indicated by circles displayed on a flattened sphere. Our $4$ training views (a-d) and $4$ validation views (e-h) are displayed in yellow and pink circles, respectively. The light yellow area represents the limited angle range of $60^\circ$, from which we obtain training views beyond the $4$ projection setting. }
    \label{fig:views}
\end{figure}

\remove{As discussed in \mbox{Section~\ref{sec:cag}}, the viewing angles in CA are literature-based optimal angles to minimize overlap and \mbox{foreshortening~\cite{di2005coronary}}.}
\addition{In clinical practice,} the LCA is typically visualized with 8 \addition{literature-based} distinctive angiographic views~\cite{green2005angiographic, green2016optimal} \addition{to minimize overlap and foreshortening~\cite{di2005coronary}}.
Figure~\ref{fig:views} shows these views in Euler angles ($\theta$, $\phi$) distributed along a flattened sphere.
The image examples are from the XCAT dataset.
For 4D reconstruction, selecting viewing angles in a sparse setting is crucial.
We, therefore, base our training and validation sets on these \addition{clinical} optimal views.
We will train our method using $4$ to $70$ projections.
For $4$ projections, we choose $4$ well-distributed optimal angles, indicated by the yellow circles in the figure and images (a-d).
The remaining views, indicated by pink circles and images (e-h), are used to validate all cases.
For more projections, we uniformly sample training views within a limited range of $60^\circ$, indicated by the light yellow area.
This area covers the range of most optimal viewing angles while avoiding views too close to the validation set. 

\subsection{Evaluation set-up} \label{subsec:evaluation}
We perform both qualitative and quantitative evaluations in our work.
The qualitative approach is suitable to overcome known problems in assessing medical imaging with pure quantitative metrics \cite{kastryulin2023image}.
For our application, where topological accuracy is most relevant, this is especially \remove{relevant}\addition{valuable}~\cite{maas2023nerf}.
We evaluate quantitatively by computing the re-projection error for 2D novel view synthesis, fitting the roadmap application where the 4D reconstruction would be re-projected to the C-arm angle of interest\addition{, as discussed in Section~\ref{sec:cag}}.

The ultimate goal of our work is to evaluate the reconstruction of the coronary artery structure rather than direct volume-to-volume comparison.
Specifically, we are interested in the topology of the structure, for which, to our knowledge, no appropriate quantitative measures have been proposed \cite{maas2023nerf}.
Therefore, we mostly rely on qualitative analysis for the structure itself.
Besides, we also perform binary comparisons by computing Dice, a well-known metric for blood vessel segmentation tasks \cite{moccia2018blood}.
We only consider the positive or blood vessel structure class when computing Dice.
This approach compensates for the imbalance of these classes due to the sparsity of the structure in the image.
To compute a binary Dice value, we generate novel views by applying maximum intensity projection (MIP) from the dynamic attenuation values $\sigma_d$, assuming that the blood vessel structure has the highest attenuation values and is purely dynamic.
This means that we strictly consider the dynamic coronary artery component for our Dice computation.
We obtain the final binarized images from thresholding the MIP views based on the 4D ground-truth values of the coronary artery structure.
When comparing to static models, we compute the MIP from the \remove{static}\addition{composite} $\sigma$ attenuation values.
We also evaluate the overall 2D CA image prediction with common image quality metrics peak-signal-to-noise ratio (PSNR) and structural similarity (SSIM). 
Besides quality metrics, we also report the computation time.

\begin{figure*}[b!]
    \centering
    \includegraphics[width=\linewidth]{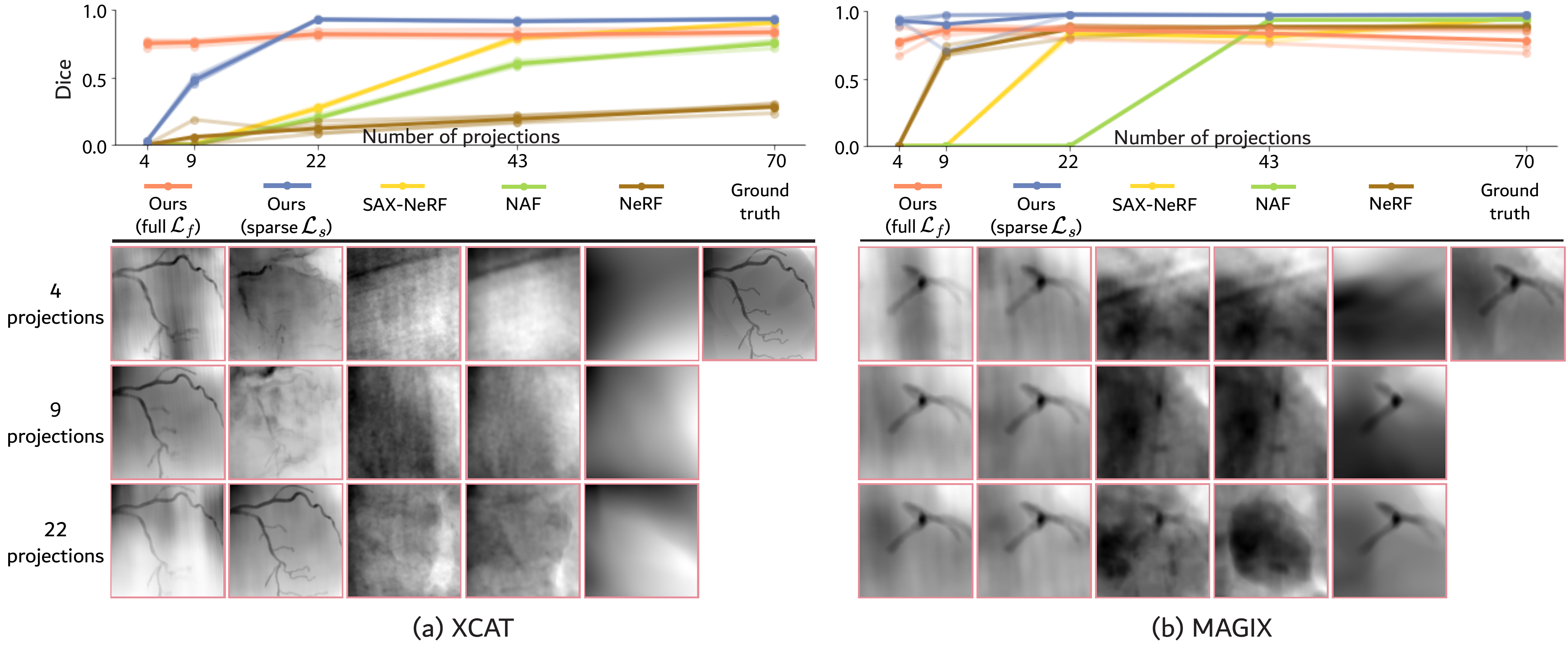}
    \caption{Dice scores and qualitative results for the comparison to existing methods across varying number of training projections, including Ours (full), Ours (sparse), SAX-NeRF \cite{cai2024structure}, NAF \cite{zha2022naf}, and NeRF \cite{mildenhall2021nerf} for the (a) XCAT and (b) MAGIX datasets. The low-opacity lines represent the scores of the four validation views, while the bold line shows the mean score across these views.}
    \label{fig:compare}
\end{figure*}

\subsection{Implementation Details}
The models were implemented in Pytorch and trained on an RTX A5000 GPU.
We utilize the same hyperparameters for both datasets.
The models have an MLP architecture of $4$ hidden layers with $128$ neurons with ReLU activations, commonly used in NeRF applications \cite{mildenhall2021nerf}.
We utilize a Softplus activation to obtain a positive attenuation value output.
We train our models with an Adam optimizer with a linearly decaying learning rate from $1 \times 10^{-3}$ to $1 \times 10^{-5}$ in $150000$ steps.
The number of training iterations is $200000$ with a ray batch size of $1024$.
We train with images of size $200 \times 200$ pixels and sample $500$ points along each ray.

The hyperparameters for our method components were chosen based on a grid search. 
Although many different hyperparameters are involved, we show they can be consistently used across different datasets\addition{, highlighting the robustness of our method}.
We impose a maximum frequency band for the windowed positional encoding, as shown in Equation~\ref{eq:pos_enc}, of $L = 12$ over $N = 150000$ iterations.
We start the coarse-to-fine transition at a frequency band of $1$, therefore activating the first frequency bands $(\sin(\pi\boldsymbol{x}), \cos(\pi\boldsymbol{x}))$ from the start of training. 
This has shown to be more advantageous for separating foreground and background.
A fraction of $V = 0.5$ of the ray batch size is sampled for the purpose of weighted pixel sampling.
The static vs. dynamic factorization loss weight $\lambda_b$ increases from $1 \times 10^{-12}$ to $1 \times 10^{-10}$, with a delay of $40000$ iterations.
The dynamic entropy loss weight $\lambda_e$ linearly increases from $1 \times 10^{-12}$ to $1 \times 10^{-10}$ without a delay, where only rays are considered with a minimum accumulated density of $1 \times 10^{-4}$.
The dynamic occlusion loss weight $\lambda_o$ increases from $1 \times 10^{-8}$ to $1 \times 10^{-5}$ linearly, delayed by $40000$ iterations with a distance threshold of $D = 0.2$.

\subsection{Comparison to Existing Methods} \label{subsec:comparison}
In this section, we compare our method to state-of-the-art 3D reconstruction methods.
As discussed in Section~\ref{subsec:works_cag}, coronary X-ray angiography techniques require manual interactions or large training datasets.
As our method does not require these settings, we compare it to state-of-the-art NeRF X-ray reconstruction methods, which obtain static reconstructions from one dataset directly.
We particularly focus on state-of-the-art methods for sparse-view reconstruction, according to the main goal of our work.
We compare our method to SAX-NeRF \cite{cai2024structure} and NAF \cite{zha2022naf}, as well as the original NeRF method as a baseline \cite{mildenhall2021nerf}.
Lastly, we include our method with only the sparse-view reconstruction components, referred to as sparse method $\mathcal{L}_s$, as a representative of a state-of-the-art sparse-view natural scene NeRF-based reconstruction method \cite{yang2023freenerf}.
This sparse method is represented by \remove{the}\addition{a part of our full loss $\mathcal{L}_f$ (see Equation~\ref{eq:full_loss}) as} loss $\mathcal{L}_s$, where $\mathcal{L}_s(\boldsymbol{r}, \boldsymbol{\tau}_i) = \mathcal{L}_p(\boldsymbol{r}, \boldsymbol{\tau}_i)$.
From our proposed components, only the photometric loss $\mathcal{L}_p$ (see Equation~\ref{eq:loss_intensity}) and the windowed positional encoding $\gamma$ (see Equation~\ref{eq:pos_enc}) is applied.
We utilize one static MLP trained individually for every single timestep.
We note that, unlike the natural scene method \cite{yang2023freenerf}, we do not apply the dynamic occlusion loss $\mathcal{L}_o$ (see Equation~\ref{eq:d_occlusion}).
This is because the assumption of edge emptiness, as described in Section~\ref{subsec:dynamicrecon}, does not apply to our static non-empty scene.

As our work focuses on the reconstruction of the coronary artery structure, we report the Dice score for this evaluation.
We refer to Section~\ref{subsec:evaluation} for details on the Dice score computation from the predicted 4D attenuation volumes.
Figure~\ref{fig:compare} shows the Dice scores for the (a) XCAT and (b) MAGIX dataset for our full method $\mathcal{L}_f$, our sparse method $\mathcal{L}_s$, SAX-NeRF \cite{cai2024structure}, NAF \cite{zha2022naf}, and NeRF \cite{mildenhall2021nerf}.
The low-opacity lines represent the scores of the four validation views, introduced in Figure~\ref{fig:views}, while the bold line shows the mean score across these views.
The figure also displays qualitative examples for the settings of $4$, $9$, and $22$ projections.
SAX-NeRF, NAF, and NeRF both show low Dice scores in the setting of $4$ and $9$ projections for both datasets.
This low score is consistent across the four validation views, as can be seen by the consistent low-opacity lines.
In the qualitative examples of these methods, we can see that the coronary artery structure is either missing or obstructed by background noise.
In this setting, our sparse method $\mathcal{L}_s$ performs well on the MAGIX dataset but fails to reconstruct a topologically accurate structure for the XCAT dataset.
Notably, our full method $\mathcal{L}_f$ significantly outperforms the other methods in this setting.
It reconstructs topologically accurate structures even with as few as $4$ training projections.
Only starting from $22$ and $43$ for the MAGIX and XCAT datasets, respectively, SAX-NeRF, NAF, and NeRF start to produce accurate reconstructions.
Although our full method $\mathcal{L}_f$ loses slight quality compared to these methods, our sparse method $\mathcal{L}_s$ still significantly outperforms them in this setting.
All in all, we significantly outperform state-of-the-art sparse-view NeRF reconstruction methods in the CA domain, with our full method $\mathcal{L}_f$ in the extremely sparse setting and our sparse method $\mathcal{L}_s$ beyond this setting.

\subsection{Ablation Studies} \label{subsec:ablation}
We evaluate the multiple components and parameters of our work through ablation studies. 
Firstly, we examine the dynamic and sparse-view reconstruction components individually, as introduced in Figure~\ref{fig:overview}~(a). 
Next, we analyze the consistency in quality for the dynamic reconstruction across the cardiac phases.
Lastly, we evaluate the effect of weighted pixel sampling and the dynamic entropy and occlusion loss components. 
\removetwo{
Furthermore, we validate our choice for self-supervision over a segmentation-supervised approach, demonstrating that the self-supervised approach achieves higher accuracy, which can be found in the supplementary material.}
\additiontwo{
Furthermore, we validate our choice for self-supervision over a segmentation-supervised approach, which can be found in the supplementary material.
Specifically, we demonstrate that the self-supervised approach achieves higher accuracy, as it does not depend on the automatic error-prone segmentation methods that produce inconsistencies in supervision. 
}\\

\begin{figure}
    \includegraphics[width=\linewidth]{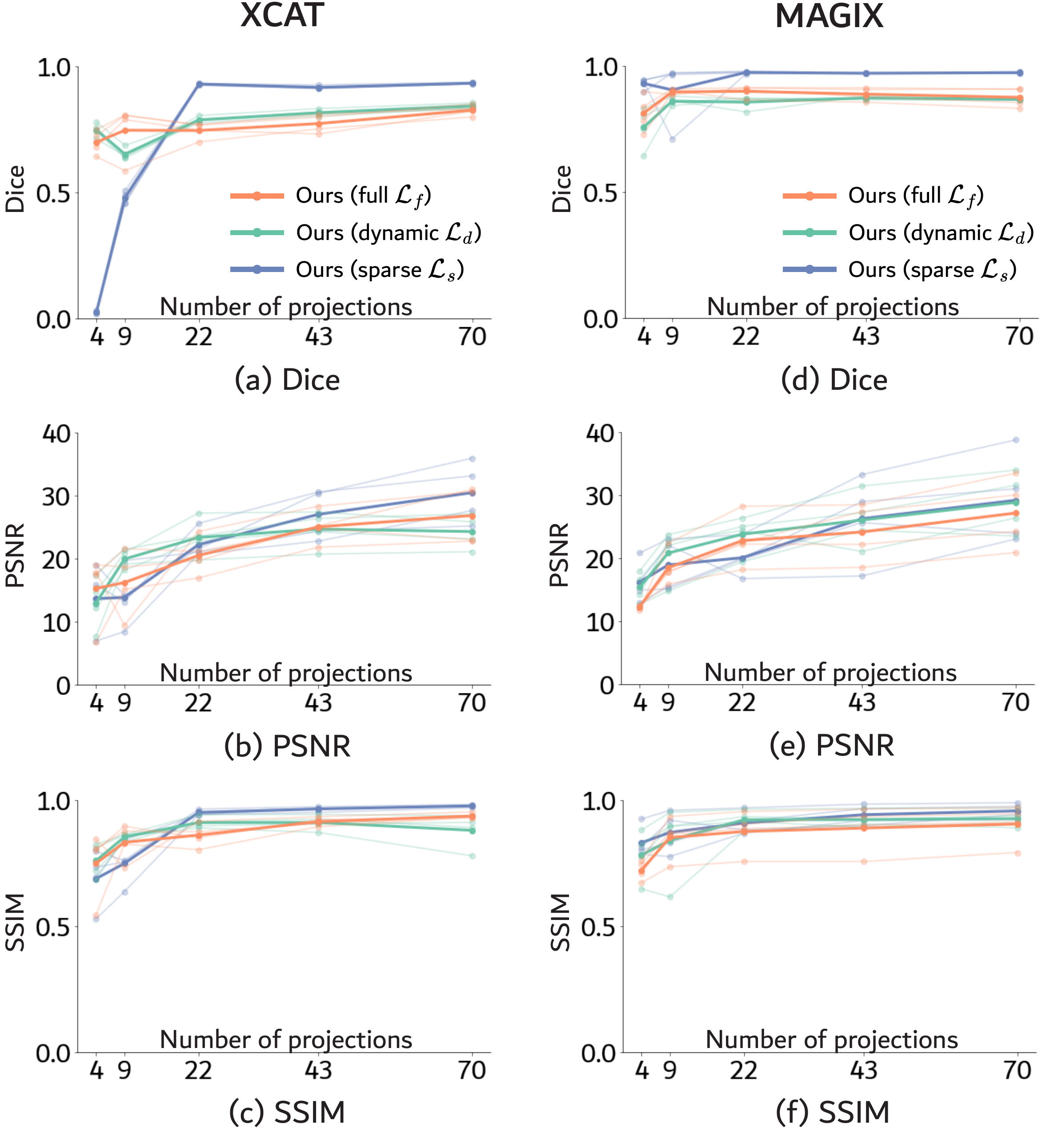}
    \caption{Quantitative results in Dice, PSNR, and SSIM for the XCAT and MAGIX datasets across different numbers of training projections. XCAT: (a) Dice, (b) PSNR, and (c) SSIM. MAGIX: (d) Dice, (e) PSNR, and (f) SSIM. We display measures for our full $\mathcal{L}_f$, dynamic $\mathcal{L}_d$, and sparse $\mathcal{L}_s$ methods. The low-opacity lines represent the scores of the four validation views, while the bold line shows the mean score across these views. }
    \label{fig:mainxcat_quant}
\end{figure}

\noindent \textbf{Dynamic and Sparse-view Reconstruction Components} \hspace{4pt} In Section~\ref{sec:method} and Figure~\ref{fig:overview}~(a), we illustrate how NeRF-CA effectively combines dynamic and sparse-view reconstruction components.
In this section, we discuss the outcomes of our full method $\mathcal{L}_f$ compared to these separate components, denoted as our sparse method and our dynamic method, respectively.
The definition of our sparse method with loss $\mathcal{L}_s$ is given in Section~\ref{subsec:comparison}.
We define our dynamic method, as shown in Figure~\ref{fig:overview}~(a) in green, \remove{with}\addition{as a part of our full loss $\mathcal{L}_f$ (see Equation~\ref{eq:full_loss}) as} loss $\mathcal{L}_d$\remove{ as} 
$$\mathcal{L}_d(\boldsymbol{r}, \boldsymbol{\tau}_i) = \mathcal{L}_p(\boldsymbol{r}, \boldsymbol{\tau}_i) + \lambda_b\mathcal{L}_b(\boldsymbol{r}, \boldsymbol{\tau}_i) \\+ \lambda_e\mathcal{L}_e(\boldsymbol{r}, \boldsymbol{\tau}_i), $$ where we also apply the windowed positional encoding $
\gamma$ (see Equation~\ref{eq:pos_enc}).
We exclude the dynamic occlusion loss $\mathcal{L}_o$ \addition{(see Equation~\ref{eq:full_loss})} to verify its role in sparse-view reconstruction. 
Figure~\ref{fig:mainxcat_quant} shows the quantitative results for the XCAT and MAGIX dataset, where (a-c) show the Dice, PSNR, and SSIM scores for the XCAT dataset and (d-f) show these scores for the MAGIX dataset.
We display these scores across various projection counts for our full $\mathcal{L}_f$ \addition{(in orange)}, sparse $\mathcal{L}_s$ \addition{(in blue)}, and dynamic $\mathcal{L}_d$ \addition{(in green)} methods.
The low-opacity lines represent the scores of the four validation views averaged over all timesteps, while the bold line shows the mean score across these views.

\begin{figure}[t]
    \centering
    \includegraphics[width=\linewidth]{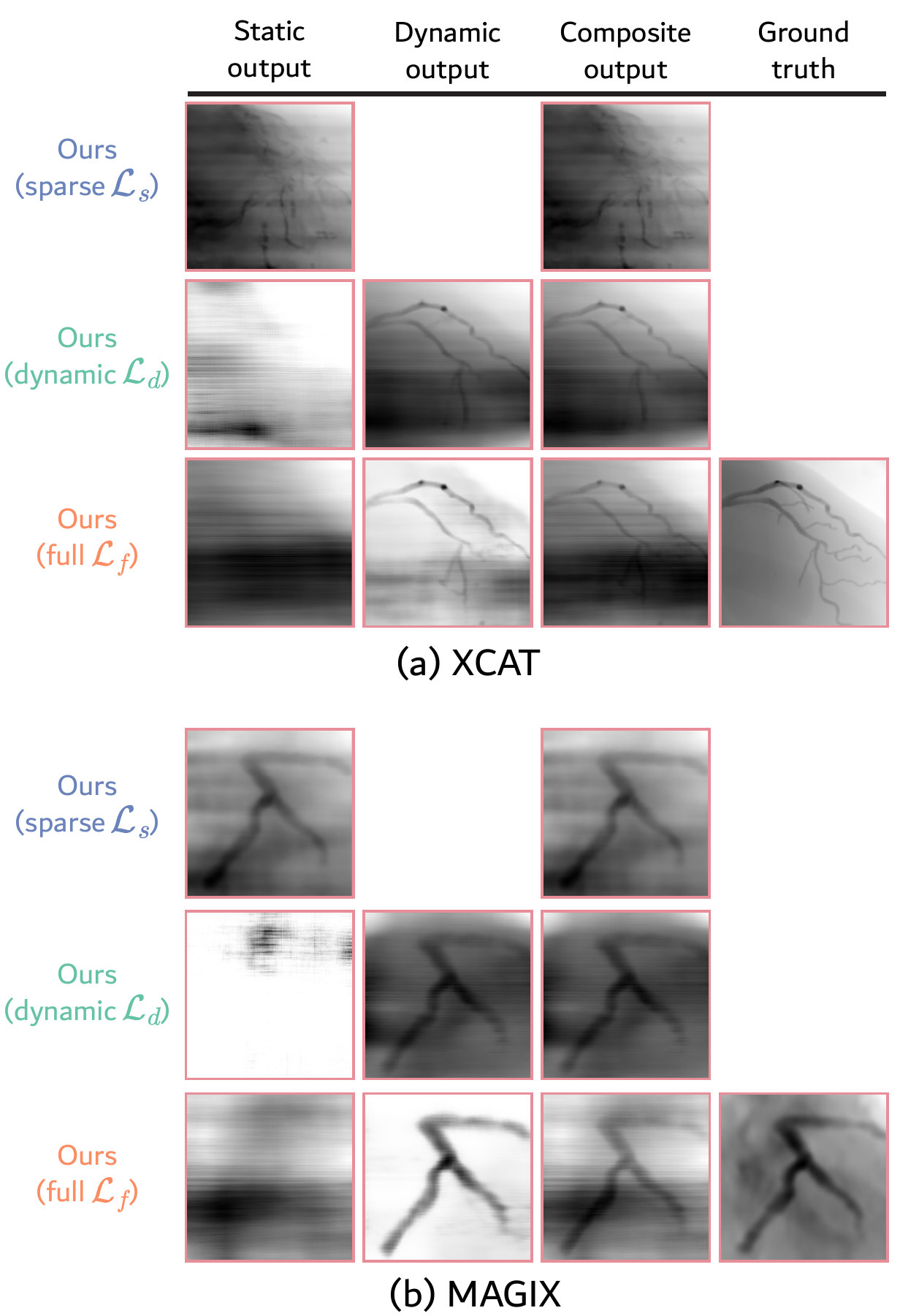}
    \caption{Qualitative results for the (a) XCAT and (b) MAGIX datasets for our full $\mathcal{L}_f$, dynamic $\mathcal{L}_d$, and sparse $\mathcal{L}_s$ methods for the setting of $4$ training projections. We show the static, dynamic, and composite output, as well as the ground truth projection.}
    \label{fig:main_qual}
\end{figure}

First, we evaluate the reconstruction of only the coronary artery structure using Dice scores, as it addresses the main goal of our work.
Our full method $\mathcal{L}_f$ \addition{(orange)} achieves high Dice scores in the extremely sparse-view settings of 4 and 9 projections.
These scores are consistent across the four validation views, as shown by the low-opacity lines.
For the XCAT dataset, our full method $\mathcal{L}_f$ significantly outperforms our sparse method $\mathcal{L}_s$ \addition{(blue)} and slightly outperforms our dynamic method $\mathcal{L}_d$ \addition{(green)}\addition{, as can be seen from Figure~\ref{fig:mainxcat_quant}~(a)}.
The scores across the MAGIX dataset are more similar\addition{, as seen in Figure~\ref{fig:mainxcat_quant}~(d)}, showing that all three of our methods perform well on this dataset, although the sparse method $\mathcal{L}_s$ slightly outperforms our other methods.
Figure~\ref{fig:main_qual} shows the qualitative examples for the $4$ training projections setting.
We show the static output, dynamic output, composite output, and ground truth images.
\remove{From these qualitative examples,}\addition{By comparing the different methods across the rows,} we can observe that our full method $\mathcal{L}_f$ \addition{(bottom row, orange)} achieves the best separation of the dynamic coronary artery from the static background, removing the occluding background from the novel views.
For the XCAT dataset~\addition{(Figure~\ref{fig:main_qual}~(a))}, the reconstructed structure is topologically accurate, with the primary branches clearly defined, which fits our clinical applications.
For the MAGIX dataset~\addition{(Figure~\ref{fig:main_qual}~(b))}, our models achieve a high-fidelity structure without loss of quality, likely due to the easier coarser nature of the structure.
We also report scores beyond the extremely sparse-view setting\addition{, as seen in Figure~\ref{fig:mainxcat_quant}(a,d)}.
For many projections, our full $\mathcal{L}_f$, dynamic $\mathcal{L}_d$, and sparse $\mathcal{L}_s$ methods all perform well in reconstructing the coronary artery structure. 
However, our full $\mathcal{L}_f$ and dynamic $\mathcal{L}_d$ methods lose quality with more training projections, whereas our sparse method $\mathcal{L}_s$ does not \addition{as can be seen from the flat line}. 
Our full method $\mathcal{L}_f$ and dynamic $\mathcal{L}_d$ method impose regularization constraints on the separation of foreground and background, which may explain the score loss with more projections. 
Although our work focuses on extremely sparse-view settings, the sparse method $\mathcal{L}_s$ might be more suitable for reconstructions with more projections.

We also report the overall CA image comparison scores in PSNR and SSIM in Figure~\ref{fig:mainxcat_quant} and the overall running time.
\remove{The PSNR scores indicate that although our full model is more susceptible to background noise, the SSIM scores indicate that the overall performance is still accurate, even in the extremely sparse setting of $4$ and $9$ projections for both datasets.}
\addition{The drop in PSNR scores (Figure~\ref{fig:main_qual}~(b, e)) indicates that our full method is susceptible to background noise. 
Meanwhile, the SSIM scores \addition{(Figure~\ref{fig:main_qual}~(c, f))} indicate that the overall performance is still accurate, even in the extremely sparse setting of $4$ and $9$ projections for both datasets.}
The sparse method $\mathcal{L}_s$ \addition{(blue)} is more accurate in background reconstruction when presented with more projections, highlighting its potential outside the extremely sparse-view setting.
We do observe variation in scores among the validation views \addition{as can be seen from the low-opacity lines in Figure~\ref{fig:mainxcat_quant}~(b,c,e,f)}, which is likely due to \remove{background occlusion}\addition{poor vessel visibility}, which can be observed in the qualitative examples \addition{in Figure~\ref{fig:main_qual}}.
However, separating the structure from the background mitigates this issue.
Therefore, our full method $\mathcal{L}_f$ remains effective, whereas our sparse $\mathcal{L}_s$ and dynamic $\mathcal{L}_d$ methods may be less applicable due to \remove{background occlusion}\addition{poor vessel visibility}.
\remove{Since the Dice score best represents our expected results and addresses this separation}\addition{Since the Dice score best reflects our application, which is reconstructing the blood vessel structure}, we will only report this score for the remaining cases.
The running time of our full method $\mathcal{L}_f$, dynamic method $\mathcal{L}_d$, and sparse method $\mathcal{L}_f$ are $5.5$, $5.5$ and $3$ hours, respectively.

Overall, our full method $\mathcal{L}_f$ effectively decouples the scene into a dynamic coronary artery structure and a static background, performing well in extremely sparse-view settings and showing high potential for coronary angiography applications.
Beyond the extremely sparse-view setting, our sparse method $\mathcal{L}_s$ outperforms our full method $\mathcal{L}_f$, highlighting its possible applications when more training projections are available. \\

\noindent \textbf{Dynamic Scene Reconstruction} \hspace{4pt} In our work, we decouple the scene into a static and dynamic component while imposing several regularizers to force the dynamic component to solely model the coronary artery structure.
We, therefore, assume that the CA scene is mostly static aside from the coronary artery structure.
As a result, any remaining background dynamics, such as contrast inhomogeneity in the MAGIX CCTA dataset, will be modeled as static, which may lead to background noise and inconsistency in output across the cardiac phases.
In this section, we evaluate whether our full method $\mathcal{L}_f$ maintains quality across the dynamic cardiac phases.

\begin{figure*}
    \centering
    \includegraphics[width=0.45\linewidth]{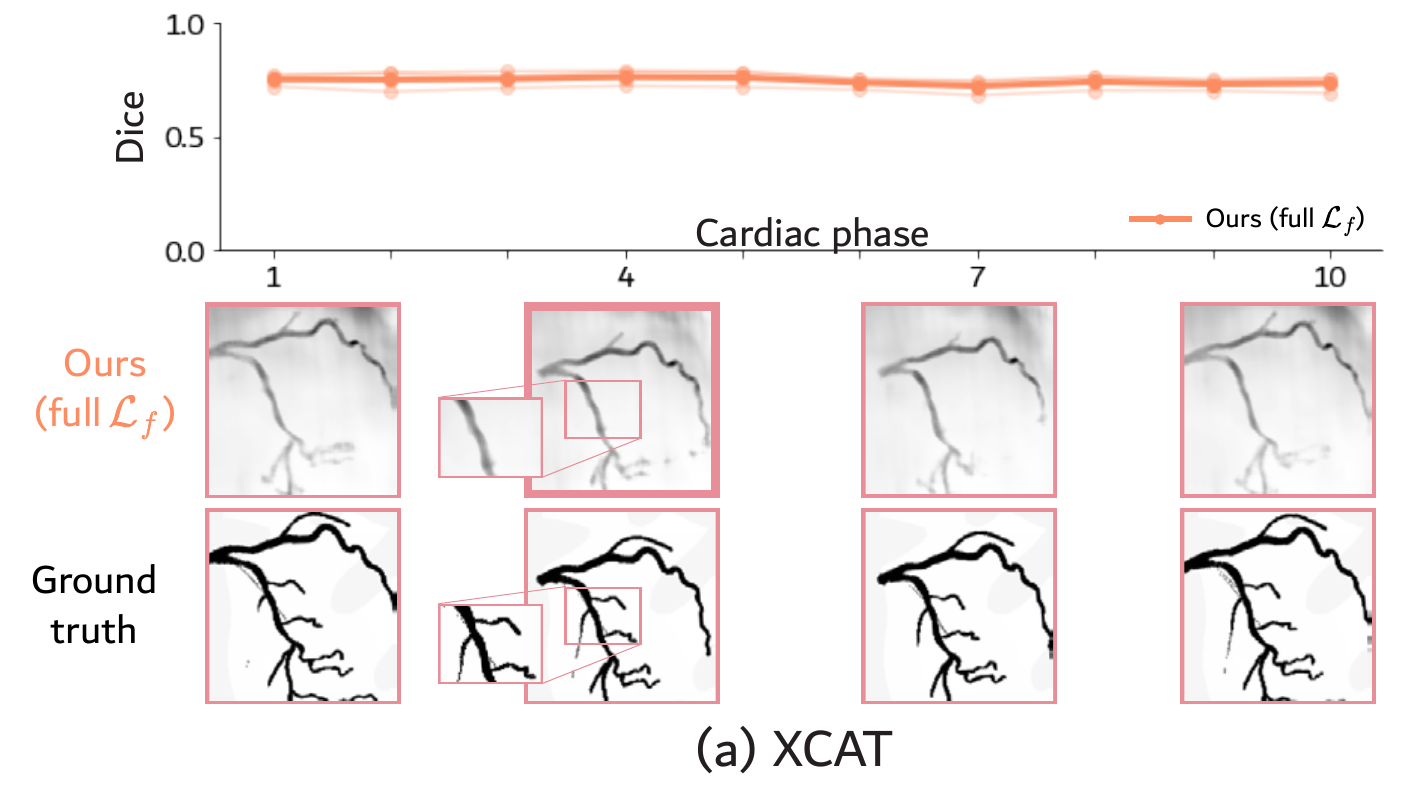}
    \hspace{0.05\linewidth}
    \includegraphics[width=0.45\linewidth]{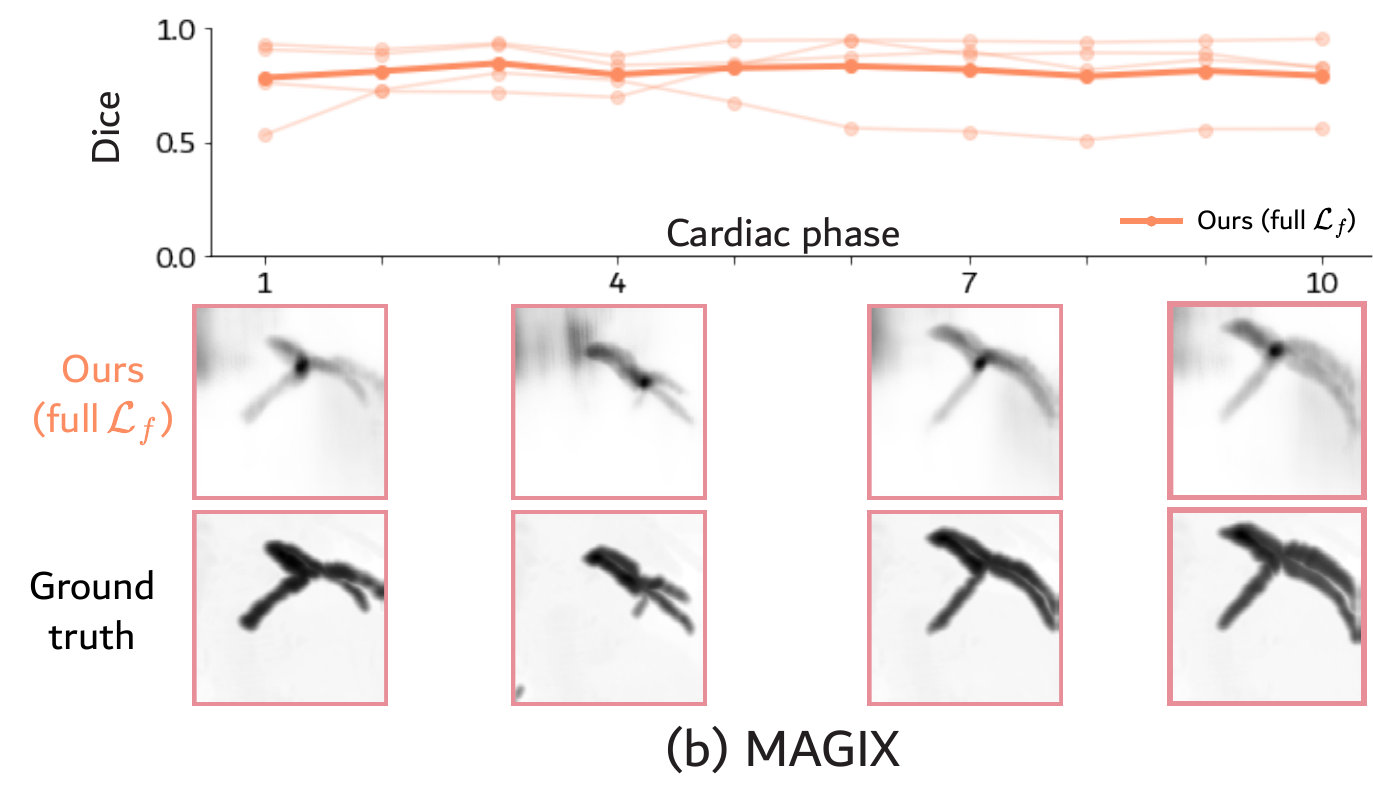}
    \caption{Dice scores for our full method $\mathcal{L}_f$ over the cardiac phases for the (a) XCAT and (b) MAGIX datasets. The model is trained with $4$ angiogram \remove{sequences}\addition{views}. The low-opacity lines represent the scores of the four validation views, whereas the bold line represents the mean score across these views. }
    \label{fig:valdyn}
\end{figure*}

Figure~\ref{fig:valdyn} shows how our method compares to the ground truth for our two datasets, (a) XCAT and (b) MAGIX, in a setting of $4$ angiogram \remove{sequences}\addition{views}.
We display the scores for the four validation views as low opacity lines, and the mean score computed from these views as a bold line.
For the (a) XCAT dataset, the line graphs indicate that the scores of our full $\mathcal{L}_f$ method are consistent across the cardiac phases, which can be derived from the constant slope of the line.
The qualitative examples, displaying the dynamic output and MIP of the ground truth, show that the slightly lower quality of the full method $\mathcal{L}_f$ is due to the missing sparse branches.
These sparse branches are less important in our application, where topology is most relevant.
Therefore, our full method $\mathcal{L}_f$ maintains quality across the dynamic cardiac phases.
For the (b) MAGIX dataset, we observe more variation in Dice score across timesteps and among the different validation views.
We can observe that this variation is due to background noise appearing in the dynamic scene.
The MAGIX dataset consists of dynamic contrast flow across the cardiac phases, resulting in these parts being modeled as dynamic by our model.
However, the background noise is primarily outside of the coronary artery structure, not affecting the reconstruction quality of the structure itself.
All in all, our model consistently maintains quality for reconstructing the coronary artery structure, fitting our clinical application. \\

\begin{figure}[b]
    \includegraphics[width=\linewidth]{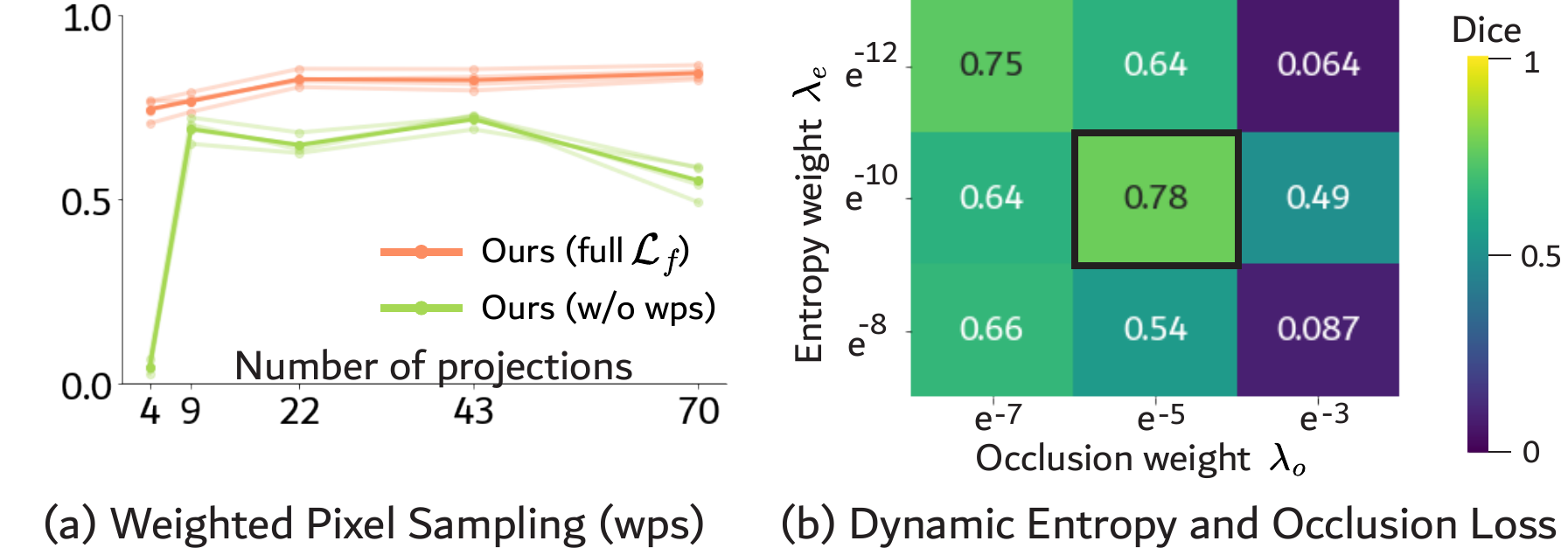}
    \caption{Dice scores for the XCAT dataset for the (a) Weighted Pixel Sampling (wps) and (b) Dynamic Entropy and Occlusion Loss ablation studies. (a) Scores for our full method (full $\mathcal{L}_f$) and our method without weighted pixel sampling (w/o wps). (b) Scores across hyperparameter combinations of the dynamic entropy weight $\lambda_e$ and dynamic occlusion weight $\lambda_o$, where the black box indicates our hyperparameter setting. }
    \label{fig:ablation_extra}
\end{figure}

\noindent \textbf{Weighted Pixel Sampling} \hspace{4pt }We incorporate weighted pixel sampling in our method to enhance the reconstruction of blood vessel structures.
This approach targets sampling in the heuristically determined regions of the blood vessels, enhancing the reconstruction of these sparse structures, as described in Section~\ref{subsec:dynamicrecon}.
Figure~\ref{fig:ablation_extra}~(a) displays the Dice scores for our full method $\mathcal{L}_f$ and our method without weighted pixel sampling (w/o wps) over the number of training projections for the XCAT dataset.
The results indicate that weighted pixel sampling \addition{(in orange in Figure~\ref{fig:ablation_extra}~(a))} generally improves the score across all projection counts, significantly boosting performance in the extremely sparse setting of $4$ projections from a nearly zero score to a high Dice score with accurate structure reconstruction.
For the MAGIX dataset, we also observe a slight overall score increase with weighted pixel sampling, as shown in the supplementary material. \\

\noindent \textbf{Dynamic Entropy and Occlusion Loss} \hspace{4pt} 
\remove{We the dynamic entropy $\mathcal{L}_e$ and dynamic occlusion  $\mathcal{L}_o$ losses to accurately decouple the scene in a static background and dynamic coronary artery, as described in \mbox{Section~\ref{subsec:dynamicrecon}} and \mbox{Section~\ref{subsec:sparserecon}}, respectively.}
\addition{In this section, we evaluate the the dynamic entropy $\mathcal{L}_e$ and dynamic occlusion  $\mathcal{L}_o$ losses, which are described in Section~\ref{subsec:dynamicrecon} and Section~\ref{subsec:sparserecon}, respectively.}
Since both losses affect the dynamic model, we analyze their combined effectiveness, as they are likely interrelated.
Figure~\ref{fig:ablation_extra}~(b) shows the Dice scores for the XCAT dataset across the dynamic entropy $\lambda_e$ and dynamic occlusion $\lambda_o$ weights for the 4 training projection settings.
The optimal weights for our model are indicated by the black square with an entropy weight $\lambda_e$ of $e^{-10}$ and the occlusion weight $\lambda_o$ of $e^{-5}$, leading to a Dice score of $0.78$.
We observe that the most optimal Dice score is obtained by combining both losses, demonstrating \remove{their effectiveness}\addition{that both are essential for optimal results}.
For the MAGIX dataset, we observe similar patterns for the same hyperparameters, as shown in the supplementary material.

\section{Discussion and Conclusion}

We propose NeRF-CA, the first step toward a 4D Neural Radiance Field reconstruction method for X-ray coronary angiography in an extremely sparse-view setting.
\addition{Our main contribution is the methodological aspects to solve the components of cardiac motion and sparse-views of the complex CA 4D reconstruction problem, given that literature has yet to propose a solution to the full clinical problem.}
\remove{W}\addition{Specifically, w}e propose a method that uniquely integrates scene decomposition with tailored regularization techniques to accurately decouple the CA scene in a static background and dynamic coronary artery structure.
We demonstrate the effectiveness of our method both in the dynamic and sparse-view reconstruction settings.
\remove{We introduce techniques such as static vs. dynamic factorization and dynamic entropy regularizers, weighted pixel sampling, and windowed positional encoding to ensure the separation of foreground and background.
We demonstrate the effectiveness of windowed positional encoding and dynamic occlusion regularization to accurately reconstruct in the extremely sparse-view setting.}
We have compared NeRF-CA to existing NeRF-based techniques for sparse-view 3D reconstruction, significantly outperforming them for CA blood vessel reconstruction.
Our ablation studies have also demonstrated the \remove{effectiveness of}\addition{need for} the individual components proposed in our method.
Notably, we achieve useful 4D reconstructions with four \addition{synthetic} angiogram \remove{sequences}\addition{views}, \addition{which are designed to capture realistic characteristics of CA data,}\remove{aligning well with clinical workflows and} showing potential for clinical adaptation.

Despite these advantages, we also highlight a few limitations of NeRF-CA.
\remove{Although showing its potential,} NeRF-CA does not \remove{fully} address all challenges in a clinical setting\remove{, such as motion or geometry inaccuracy in CA data}.
Motion-related challenges \addition{that are not addressed} include respiratory motion, contrast inhomogeneity, and synchronization across viewpoints.
Specifically, our method assumes synchronous motion with the ECG signal across viewpoints, which \remove{may no longer hold when incorporating these additional challenges.}\addition{cannot be assumed in clinical data due to these additional challenges.}
\remove{Therefore, we rely on synthetic CA data to perform quantitative evaluation.}
\remove{Future work should consider the various CA-based works proposed to deal with these \mbox{challenges~\cite{ccimen2016reconstruction}}, although modifications may be needed to incorporate them into NeRF-CA.}
\addition{Future work could take inspiration from the various CA-based works that propose specific techniques to deal with these motion-related challenges in CA data~\cite{ccimen2016reconstruction}.
Nevertheless, integrating these techniques into NeRF-CA is complex, as they typically rely on deformable models~\cite{baka2015respiratory} or image registration~\cite{chen2009three}, which require human interactions, such as segmentations. 
In contrast, NeRF-CA implicitly models motion without segmentations, making direct integration non-trivial and beyond the scope of this work.}
\addition{Furthermore, we validate our method using synthetic data, which has limitations such as simplified motion and homogeneous vessel contrast.
However, this approach enabled a quantitative evaluation of NeRF-CA and its components.
Given a further developed method which addresses all challenges in real CA data, testing on real data would be a necessary step.}
Additionally, geometry inaccuracies from C-arm calibration in clinical settings, unlike the perfect camera geometry assumed in our work, could impact performance and require further investigation.
Finally, improving the running time of our method from hours to minutes is crucial for integration into clinical workflows. 
Future work could explore techniques like volumetric \mbox{representations~\cite{zha2022naf}} or \mbox{3D Gaussian Splatting~\cite{cai2024radiative}}, which have shown promise in accelerating NeRF-based medical reconstructions.
\remove{However, integrating these techniques requires significant adaptations and is therefore beyond the scope of this work. }
\addition{However, integrating these techniques is not straightforward, given that these methods do not represent the scene as a continuous space, which we rely on in our approach, both for static-dynamic decomposition and sparse-view modelling.
Therefore, these existing frameworks would require large adaptations and are beyond the scope of this work.}

In this work, we proposed NeRF-CA, the first step toward a 4D reconstruction method for CA data in the extremely sparse-view setting without user interaction.
We demonstrate \remove{accurate}\addition{adequate} blood vessel reconstructions from as few as four angiogram \remove{sequences}\addition{views}\remove{, matching the clinical workflows}\addition{with synthetic data, showing the potential for clinical applications} and outperforming state-of-the-art NeRF-based 3D reconstruction techniques.
\remove{Future work includes applying our work on clinical data and reducing computation time.}

\section*{Acknowledgements}
This research was performed within the MEDUSA project as part of
the Eindhoven MedTech Innovation Center collaboration
between Eindhoven University of Technology, Philips Medical Systems, and Catharina Ziekenhuis Eindhoven.

\bibliographystyle{IEEEtran}
\bibliography{references}

\begin{thebibliography}{10}
\providecommand{\url}[1]{#1}
\csname url@samestyle\endcsname
\providecommand{\newblock}{\relax}
\providecommand{\bibinfo}[2]{#2}
\providecommand{\BIBentrySTDinterwordspacing}{\spaceskip=0pt\relax}
\providecommand{\BIBentryALTinterwordstretchfactor}{4}
\providecommand{\BIBentryALTinterwordspacing}{\spaceskip=\fontdimen2\font plus
\BIBentryALTinterwordstretchfactor\fontdimen3\font minus \fontdimen4\font\relax}
\providecommand{\BIBforeignlanguage}[2]{{%
\expandafter\ifx\csname l@#1\endcsname\relax
\typeout{** WARNING: IEEEtran.bst: No hyphenation pattern has been}%
\typeout{** loaded for the language `#1'. Using the pattern for}%
\typeout{** the default language instead.}%
\else
\language=\csname l@#1\endcsname
\fi
#2}}
\providecommand{\BIBdecl}{\relax}
\BIBdecl

\bibitem{green2004three}
N.~E. Green, S.-Y.~J. Chen, J.~C. Messenger, B.~M. Groves, and J.~D. Carroll, ``Three-dimensional vascular angiography,'' \emph{Current problems in cardiology}, vol.~29, no.~3, pp. 104--142, 2004.

\bibitem{piayda2018dynamic}
K.~Piayda, L.~Kleinebrecht, S.~Afzal, R.~Bullens, I.~Ter~Horst, A.~Polzin, V.~Veulemans, L.~Dannenberg, A.~C. Wimmer, C.~Jung \emph{et~al.}, ``Dynamic coronary roadmapping during percutaneous coronary intervention: a feasibility study,'' \emph{European journal of medical research}, vol.~23, pp. 1--7, 2018.

\bibitem{ccimen2016reconstruction}
S.~{\c{C}}imen, A.~Gooya, M.~Grass, and A.~F. Frangi, ``Reconstruction of coronary arteries from x-ray angiography: A review,'' \emph{Medical image analysis}, vol.~32, pp. 46--68, 2016.

\bibitem{iyer2023multi}
K.~Iyer, B.~K. Nallamothu, C.~A. Figueroa, and R.~R. Nadakuditi, ``A multi-stage neural network approach for coronary 3d reconstruction from uncalibrated x-ray angiography images,'' \emph{Scientific Reports}, vol.~13, no.~1, p. 17603, 2023.

\bibitem{zhu2025sparse}
Y.~Zhu, Y.~Wang, C.~Di, H.~Liu, F.~Liao, and S.~Ma, ``Sparse and transferable three-dimensional dynamic vascular reconstruction for instantaneous diagnosis,'' \emph{Nature Machine Intelligence}, pp. 1--13, 2025.

\bibitem{mildenhall2021nerf}
B.~Mildenhall, P.~P. Srinivasan, M.~Tancik, J.~T. Barron, R.~Ramamoorthi, and R.~Ng, ``Nerf: Representing scenes as neural radiance fields for view synthesis,'' \emph{Communications of the ACM}, vol.~65, no.~1, pp. 99--106, 2021.

\bibitem{maas2023nerf}
K.~W.~H. Maas, N.~Pezzotti, A.~J.~E. Vermeer, D.~Ruijters, and A.~Vilanova, ``Nerf for 3d reconstruction from x-ray angiography: Possibilities and limitations,'' in \emph{VCBM 2023: Eurographics Workshop on Visual Computing for Biology and Medicine}.\hskip 1em plus 0.5em minus 0.4em\relax Eurographics Association, 2023, pp. 29--40.

\bibitem{gao2022nerf}
K.~Gao, Y.~Gao, H.~He, D.~Lu, L.~Xu, and J.~Li, ``Nerf: Neural radiance field in 3d vision, a comprehensive review,'' \emph{arXiv preprint arXiv:2210.00379}, 2022.

\bibitem{yang2023freenerf}
J.~Yang, M.~Pavone, and Y.~Wang, ``Freenerf: Improving few-shot neural rendering with free frequency regularization,'' in \emph{Proceedings of the IEEE/CVF Conference on Computer Vision and Pattern Recognition}, 2023, pp. 8254--8263.

\bibitem{cai2024structure}
Y.~Cai, J.~Wang, A.~Yuille, Z.~Zhou, and A.~Wang, ``Structure-aware sparse-view x-ray 3d reconstruction,'' in \emph{Proceedings of the IEEE/CVF Conference on Computer Vision and Pattern Recognition}, 2024, pp. 11\,174--11\,183.

\bibitem{fu20243dgr}
X.~Fu, Y.~Li, F.~Tang, J.~Li, M.~Zhao, G.-J. Teng, and S.~K. Zhou, ``3dgr-car: Coronary artery reconstruction from ultra-sparse 2d x-ray views with a 3d gaussians representation,'' in \emph{International Conference on Medical Image Computing and Computer-Assisted Intervention}.\hskip 1em plus 0.5em minus 0.4em\relax Springer, 2024, pp. 14--24.

\bibitem{wang2024neca}
Y.~Wang, A.~Banerjee, and V.~Grau, ``Neca: 3d coronary artery tree reconstruction from two 2d projections via neural implicit representation,'' \emph{Bioengineering}, vol.~11, no.~12, p. 1227, 2024.

\bibitem{liu20243ddsa}
Z.~Liu, H.~Zhao, W.~Qin, Z.~Zhou, X.~Wang, W.~Wang, X.~Lai, C.~Zheng, D.~Shen, and Z.~Cui, ``3d vessel reconstruction from sparse-view dynamic dsa images via vessel probability guided attenuation learning,'' \emph{arXiv preprint arXiv:2405.10705}, 2024.

\bibitem{wu2022d}
T.~Wu, F.~Zhong, A.~Tagliasacchi, F.~Cole, and C.~Oztireli, ``D\^{} 2nerf: Self-supervised decoupling of dynamic and static objects from a monocular video,'' \emph{Advances in neural information processing systems}, vol.~35, pp. 32\,653--32\,666, 2022.

\bibitem{kim2022infonerf}
M.~Kim, S.~Seo, and B.~Han, ``Infonerf: Ray entropy minimization for few-shot neural volume rendering,'' in \emph{Proceedings of the IEEE/CVF Conference on Computer Vision and Pattern Recognition}, 2022, pp. 12\,912--12\,921.

\bibitem{segars20104d}
W.~P. Segars, G.~Sturgeon, S.~Mendonca, J.~Grimes, and B.~M. Tsui, ``4d xcat phantom for multimodality imaging research,'' \emph{Medical physics}, vol.~37, no.~9, pp. 4902--4915, 2010.

\bibitem{rosset2004osirix}
A.~Rosset, L.~Spadola, and O.~Ratib, ``Osirix: an open-source software for navigating in multidimensional dicom images,'' \emph{Journal of digital imaging}, vol.~17, pp. 205--216, 2004.

\bibitem{chen2009three}
S.~J. Chen and D.~Sch{\"a}fer, ``Three-dimensional coronary visualization, part 1: modeling,'' \emph{Cardiology clinics}, vol.~27, no.~3, pp. 433--452, 2009.

\bibitem{knuuti20202019}
J.~Knuuti, W.~Wijns, A.~Saraste, D.~Capodanno, E.~Barbato, C.~Funck-Brentano, E.~Prescott, R.~F. Storey, C.~Deaton, T.~Cuisset \emph{et~al.}, ``2019 esc guidelines for the diagnosis and management of chronic coronary syndromes: The task force for the diagnosis and management of chronic coronary syndromes of the european society of cardiology (esc),'' \emph{European heart journal}, vol.~41, no.~3, pp. 407--477, 2020.

\bibitem{shechter2006displacement}
G.~Shechter, J.~R. Resar, and E.~R. McVeigh, ``Displacement and velocity of the coronary arteries: cardiac and respiratory motion,'' \emph{IEEE transactions on medical imaging}, vol.~25, no.~3, pp. 369--375, 2006.

\bibitem{tancik2020fourier}
M.~Tancik, P.~Srinivasan, B.~Mildenhall, S.~Fridovich-Keil, N.~Raghavan, U.~Singhal, R.~Ramamoorthi, J.~Barron, and R.~Ng, ``Fourier features let networks learn high frequency functions in low dimensional domains,'' \emph{Advances in neural information processing systems}, vol.~33, pp. 7537--7547, 2020.

\bibitem{max1995optical}
N.~Max, ``Optical models for direct volume rendering,'' \emph{IEEE Transactions on Visualization and Computer Graphics}, vol.~1, no.~2, pp. 99--108, 1995.

\bibitem{hwang2021simple}
M.~Hwang, S.-B. Hwang, H.~Yu, J.~Kim, D.~Kim, W.~Hong, A.-J. Ryu, H.~Y. Cho, J.~Zhang, B.~K. Koo \emph{et~al.}, ``A simple method for automatic 3d reconstruction of coronary arteries from x-ray angiography,'' \emph{Frontiers in Physiology}, vol.~12, p. 724216, 2021.

\bibitem{bappy2021automated}
D.~Bappy, A.~Hong, E.~Choi, J.-O. Park, and C.-S. Kim, ``Automated three-dimensional vessel reconstruction based on deep segmentation and bi-plane angiographic projections,'' \emph{Computerized Medical Imaging and Graphics}, vol.~92, p. 101956, 2021.

\bibitem{zhao2022self}
H.~Zhao, Z.~Zhou, F.~Wu, D.~Xiang, H.~Zhao, W.~Zhang, L.~Li, Z.~Li, J.~Huang, H.~Hu \emph{et~al.}, ``Self-supervised learning enables 3d digital subtraction angiography reconstruction from ultra-sparse 2d projection views: a multicenter study,'' \emph{Cell Reports Medicine}, vol.~3, no.~10, 2022.

\bibitem{yunus2024recent}
R.~Yunus, J.~E. Lenssen, M.~Niemeyer, Y.~Liao, C.~Rupprecht, C.~Theobalt, G.~Pons-Moll, J.-B. Huang, V.~Golyanik, and E.~Ilg, ``Recent trends in 3d reconstruction of general non-rigid scenes,'' in \emph{Computer Graphics Forum}.\hskip 1em plus 0.5em minus 0.4em\relax Wiley Online Library, 2024, p. e15062.

\bibitem{wang2024neural}
X.~Wang, S.~Hu, H.~Fan, H.~Zhu, and X.~Li, ``Neural radiance fields in medical imaging: Challenges and next steps,'' \emph{arXiv preprint arXiv:2402.17797}, 2024.

\bibitem{molaei2023implicit}
A.~Molaei, A.~Aminimehr, A.~Tavakoli, A.~Kazerouni, B.~Azad, R.~Azad, and D.~Merhof, ``Implicit neural representation in medical imaging: A comparative survey,'' in \emph{Proceedings of the IEEE/CVF International Conference on Computer Vision}, 2023, pp. 2381--2391.

\bibitem{niemeyer2022regnerf}
M.~Niemeyer, J.~T. Barron, B.~Mildenhall, M.~S. Sajjadi, A.~Geiger, and N.~Radwan, ``Regnerf: Regularizing neural radiance fields for view synthesis from sparse inputs,'' in \emph{Proceedings of the IEEE/CVF Conference on Computer Vision and Pattern Recognition}, 2022, pp. 5480--5490.

\bibitem{jain2021putting}
A.~Jain, M.~Tancik, and P.~Abbeel, ``Putting nerf on a diet: Semantically consistent few-shot view synthesis,'' in \emph{Proceedings of the IEEE/CVF International Conference on Computer Vision}, 2021, pp. 5885--5894.

\bibitem{ruckert2022neat}
D.~R{\"u}ckert, Y.~Wang, R.~Li, R.~Idoughi, and W.~Heidrich, ``Neat: Neural adaptive tomography,'' \emph{ACM Transactions on Graphics (TOG)}, vol.~41, no.~4, pp. 1--13, 2022.

\bibitem{zang2021intratomo}
G.~Zang, R.~Idoughi, R.~Li, P.~Wonka, and W.~Heidrich, ``Intratomo: self-supervised learning-based tomography via sinogram synthesis and prediction,'' in \emph{Proceedings of the IEEE/CVF International Conference on Computer Vision}, 2021, pp. 1960--1970.

\bibitem{lin2023learning}
Y.~Lin, Z.~Luo, W.~Zhao, and X.~Li, ``Learning deep intensity field for extremely sparse-view cbct reconstruction,'' in \emph{International Conference on Medical Image Computing and Computer-Assisted Intervention}.\hskip 1em plus 0.5em minus 0.4em\relax Springer, 2023, pp. 13--23.

\bibitem{fang2022snaf}
Y.~Fang, L.~Mei, C.~Li, Y.~Liu, W.~Wang, Z.~Cui, and D.~Shen, ``Snaf: Sparse-view cbct reconstruction with neural attenuation fields,'' \emph{arXiv preprint arXiv:2211.17048}, 2022.

\bibitem{kshirsagar2024generative}
J.~Kshirsagar, J.~McNulty, B.~Taji, D.~So, A.-Y. Chong, P.~Theriault-Lauzier, A.~Wisniewski, and S.~Shrimohammadi, ``Generative ai-assisted novel view synthesis of coronary arteries for angiography,'' in \emph{2024 IEEE International Symposium on Medical Measurements and Applications (MeMeA)}.\hskip 1em plus 0.5em minus 0.4em\relax IEEE, 2024, pp. 1--6.

\bibitem{zha2022naf}
R.~Zha, Y.~Zhang, and H.~Li, ``Naf: Neural attenuation fields for sparse-view cbct reconstruction,'' in \emph{International Conference on Medical Image Computing and Computer-Assisted Intervention}.\hskip 1em plus 0.5em minus 0.4em\relax Springer, 2022, pp. 442--452.

\bibitem{zhou2023tiavox}
Z.~Zhou, H.~Zhao, J.~Fang, D.~Xiang, L.~Chen, L.~Wu, F.~Wu, W.~Liu, C.~Zheng, and X.~Wang, ``Tiavox: Time-aware attenuation voxels for sparse-view 4d dsa reconstruction,'' \emph{arXiv preprint arXiv:2309.02318}, 2023.

\bibitem{liu2014improved}
B.~Liu, F.~Zhou, and X.~Bai, ``Improved c-arm cardiac cone beam ct based on alternate reconstruction and segmentation,'' \emph{Biomedical Signal Processing and Control}, vol.~13, pp. 113--122, 2014.

\bibitem{frangi1998multiscale}
A.~F. Frangi, W.~J. Niessen, K.~L. Vincken, and M.~A. Viergever, ``Multiscale vessel enhancement filtering,'' in \emph{Medical Image Computing and Computer-Assisted Intervention—MICCAI’98: First International Conference Cambridge, MA, USA, October 11--13, 1998 Proceedings 1}.\hskip 1em plus 0.5em minus 0.4em\relax Springer, 1998, pp. 130--137.

\bibitem{li2004improved}
M.~Li, H.~Kudo, J.~Hu, and R.~H. Johnson, ``Improved iterative algorithm for sparse object reconstruction and its performance evaluation with micro-ct data,'' \emph{IEEE Transactions on Nuclear Science}, vol.~51, no.~3, pp. 659--666, 2004.

\bibitem{lin2021barf}
C.-H. Lin, W.-C. Ma, A.~Torralba, and S.~Lucey, ``Barf: Bundle-adjusting neural radiance fields,'' in \emph{Proceedings of the IEEE/CVF International Conference on Computer Vision}, 2021, pp. 5741--5751.

\bibitem{park2021nerfies}
K.~Park, U.~Sinha, J.~T. Barron, S.~Bouaziz, D.~B. Goldman, S.~M. Seitz, and R.~Martin-Brualla, ``Nerfies: Deformable neural radiance fields,'' in \emph{Proceedings of the IEEE/CVF International Conference on Computer Vision}, 2021, pp. 5865--5874.

\bibitem{rohkohl2010cavarev}
C.~Rohkohl, G.~Lauritsch, A.~Keil, and J.~Hornegger, ``Cavarev—an open platform for evaluating 3d and 4d cardiac vasculature reconstruction,'' \emph{Physics in Medicine \& Biology}, vol.~55, no.~10, p. 2905, 2010.

\bibitem{biguri2016tigre}
A.~Biguri, M.~Dosanjh, S.~Hancock, and M.~Soleimani, ``Tigre: a matlab-gpu toolbox for cbct image reconstruction,'' \emph{Biomedical Physics \& Engineering Express}, vol.~2, no.~5, p. 055010, 2016.

\bibitem{green2005angiographic}
N.~E. Green, S.-Y.~J. Chen, A.~R. Hansgen, J.~C. Messenger, B.~M. Groves, and J.~D. Carroll, ``Angiographic views used for percutaneous coronary interventions: A three-dimensional analysis of physician-determined vs. computer-generated views,'' \emph{Catheterization and Cardiovascular Interventions}, vol.~64, no.~4, pp. 451--459, 2005.

\bibitem{green2016optimal}
P.~Green, P.~Frobisher, and S.~Ramcharitar, ``Optimal angiographic views for invasive coronary angiography: A guide for trainees,'' \emph{Br J Cardiol}, vol.~23, pp. 110--3, 2016.

\bibitem{di2005coronary}
C.~Di~Mario and N.~Sutaria, ``Coronary angiography in the angioplasty era: projections with a meaning,'' \emph{Heart}, vol.~91, no.~7, pp. 968--976, 2005.

\bibitem{kastryulin2023image}
S.~Kastryulin, J.~Zakirov, N.~Pezzotti, and D.~V. Dylov, ``Image quality assessment for magnetic resonance imaging,'' \emph{IEEE Access}, vol.~11, pp. 14\,154--14\,168, 2023.

\bibitem{moccia2018blood}
S.~Moccia, E.~De~Momi, S.~El~Hadji, and L.~S. Mattos, ``Blood vessel segmentation algorithms—review of methods, datasets and evaluation metrics,'' \emph{Computer methods and programs in biomedicine}, vol. 158, pp. 71--91, 2018.

\bibitem{baka2015respiratory}
N.~Baka, B.~Lelieveldt, C.~Schultz, W.~Niessen, and T.~van Walsum, ``Respiratory motion estimation in x-ray angiography for improved guidance during coronary interventions,'' \emph{Physics in Medicine \& Biology}, vol.~60, no.~9, p. 3617, 2015.

\bibitem{cai2024radiative}
Y.~Cai, Y.~Liang, J.~Wang, A.~Wang, Y.~Zhang, X.~Yang, Z.~Zhou, and A.~Yuille, ``Radiative gaussian splatting for efficient x-ray novel view synthesis,'' \emph{arXiv preprint arXiv:2403.04116}, 2024.

\end{thebibliography}

 




\end{document}